\documentclass[
twocolumn, 
amsmath,amssymb,floatfix,10pt,prd,
superscriptaddress,nofootinbib]{revtex4}

\usepackage[parfill]{parskip}    
\usepackage{graphicx}
\usepackage{amssymb}
\usepackage{epstopdf}
\usepackage{color}
\usepackage{float}
\usepackage{amsmath}
\usepackage{orcidlink}
\usepackage{latexsym}
\usepackage{amsmath}
\usepackage{amssymb}
\usepackage{amsfonts}
\usepackage{subcaption}

\begin{document}

\title{Gravitationally decoupled Non-Schwarzschild black holes and wormhole space-times }

\author{Francisco Tello-Ortiz
 \orcidlink{0000-0002-7104-5746}}
 \email{francisco.tello@pucv.cl}
\affiliation{
Instituto de Física, Pontificia Universidad Católica de Valparaíso, Casilla 4950, Valparaíso, Chile.}

\author{Ángel Rincón
\orcidlink{0000-0001-8069-9162}
}
\email{angel.rincon@ua.es}
\affiliation{Departamento de Física Aplicada, Universidad de Alicante, Campus de San Vicente del Raspeig, E-03690 Alicante, Spain.}

\author{A. Alvarez}
\email{aalvarezu88@gmail.com}
\affiliation{Unidad de Equipamiento Cient\'ifico (Maini), \\ Universidad Cat\'olica del Norte, 
Av. Angamos 0610, Antofagasta, Chile.}

\author{Saibal Ray
\orcidlink{0000-0002-5909-0544}}
\email{saibal.ray@gla.ac.in}
\affiliation{Centre for Cosmology, Astrophysics and Space Science (CCASS),
GLA University, Mathura 281406, Uttar Pradesh, India.}

\begin{abstract}
In this article, using gravitational decoupling by means of minimal geometric deformation approach, we obtain a new spherically symmetric and static black hole solution. To progress, we close the system by assuming that the average pressure of the $\theta$-sector is vanishing. Also, we tackle the problem regarding how, for a given minimally deformed black hole solution, one can connect it to a wormhole space-time.
\end{abstract}

\maketitle

\section{Introduction}

Black holes are logically one of the most attractive as well as interesting compact objects in the field of gravitation. Such interests are two folds: from a purely theoretical perspective \cite{book:249053} and from an observational point of view \cite{Bambi:2019xzp}. They essentially merge together different fields of physics, especially, thermodynamical physics as well as quantum mechanics. Therefore, starting from General Relativity (GR), passing by thermodynamics and quantum field theory, they do now also impact into the regime of high energy physics, i.e., particle and collider physics. 

Thus, black holes are interesting not only by their exotic nature, but also because they are an ideal scenario to investigate gravity in the strong field regime, testing several aspects of GR.
A few classical and remarkable examples of black holes in four dimensions are the following:
(i) the Schwarzschild solution \cite{Schwarzschild:1916uq},
(ii) the Reissner-Nordstr\"om solution \cite{1916AnP...355..106R,1918KNAB...20.1238N},
(iii) the Kerr solution \cite{Kerr:1963ud} and, finally,
(iv) the Kerr-Newman solution \cite{Newman:1965my}.

In particular, black holes are parameterized by three fundamental values: (i) the mass $M$, (ii) the angular momentum $J$ and finally, (iii) the electric charge $Q$ ~\cite{Hawking:1971vc}. Such affirmation is supported by the no-hair theorem, which claims that these solutions should not carry any other charges~\cite{Ruffini:1971bza}. Importantly, due to inner gauge symmetries, black holes can also have ``hairs'', i.e., additional charges which parameterize the black hole solution \cite{Hawking:2016msc}.

The study of black hole physics is quite robust which includes: (i) quasinormal modes (see classical reviews \cite{Kokkotas:1999bd,Berti:2009kk,Konoplya:2011qq}, and a few recent works \cite{Panotopoulos:2017hns,Li:2022kch,Rincon:2018sgd,Destounis:2018utr,Fernando:2022wlm,Gonzalez:2018xrq}), (ii) regular solutions and also solutions with non-linear electrodynamics \cite{Tzikas:2018cvs,Bargueno:2020ais,Panotopoulos:2019qjk}, (iii) geodesic analysis \cite{Cruz:2004ts,Leiva:2008fd,Panotopoulos:2022bky}, among other.
Similarly, black holes and their properties have been studied beyond Einstein's gravity. To name a few examples, we can mention $f(R),~f(G),~f(T)$ (see \cite{delaCruz-Dombriz:2009pzc,Moon:2011hq,deSousaSilva:2018kkt,Junior:2015fya}),  Brans-Dicke theory \cite{Kim:1998hc,Panotopoulos:2022xuj}, Asymptotically safe gravity \cite{Falls:2010he,Cai:2010zh}, scale-dependent gravity \cite{Koch:2016uso,Rincon:2018lyd,Rincon:2020cpz,Ovgun:2023ego}, improved black hole solutions \cite{Bonanno:2000ep} and many other approaches. 

As mentioned before, these objects seem to be characterized only by the mass $M$, the electric charge $Q$ and the angular momentum $J$. In order to bypass the no--go theorem, we shall take  advantage of a recent method that assume the existence of an additional source described by a conserved energy--momentum tensor $\theta_{\mu\nu}$. Albeit the impact of an additional source in the energy--momentum tensor has been investigated on many occasions, in this case the novel property is that the sources do not directly interact with the matter that sources the (hairless) black hole solution. This method called gravitational decoupling (GD) by minimal geometric deformation (MGD), was developed in Refs.~\cite{Ovalle:2017fgl,Ovalle:2017wqi} precisely to make progress in such direction. It is worth mentioning that, this is the first scheme applying GD in a consistent way, within the framework of GR, providing analytical solutions with physical content and interpretation. 

Basically, GD by MGD can be implemented in two basic forms: (i) extending solutions into extended domains, or (ii) deconstructing a gravitational source into simple parts.

In the following sections, we shall introduce the main idea and details behind GD by MGD.
Still, at this level, it is essential to mention that the method has been tested several times. To be more precise, new hairy black holes and some of their properties have been explored in Refs. \cite{Ovalle:2018umz,Contreras:2018nfg,Estrada:2020ptc,Fernandes-Silva:2019fez,daRocha:2020gee,Estrada:2021kuj}. Even more, within the framework of the so--called extended MGD (e--MGD) \cite{Ovalle:2018gic}, more complex hairy black holes have been found and their properties thoroughly studied in Refs. \cite{Ovalle:2020kpd,Contreras:2021yxe,Ramos:2021jta,Sultana:2021cvq,Ovalle:2021jzf,Meert:2021khi,Arias:2022jax,Cavalcanti:2022cga,Cavalcanti:2022adb,Casadio:2022ndh,Ovalle:2022eqb,Panotopoulos:2018law,Avalos:2023ywb,Avalos:2023jeh,Ovalle:2023vvu,Casadio:2023iqt}.

On the other hand, beyond the scope of black hole solutions, a plethora of articles have been published using GD by means of MGD or e--MGD, to obtain new toy models representing compact objects such as neutron stars from Einstein's gravity to modify gravity theories  \cite{Gabbanelli:2018bhs,Ovalle:2018ans,Ovalle:2019lbs,Gabbanelli:2019txr,Contreras:2019iwm,Torres-Sanchez:2019wjv,Casadio:2019usg,Abellan:2020wjw,Abellan:2020dze,Contreras:2022vec,Santana:2022vmw,Contreras:2022nji,Ovalle:2022yjl,Andrade:2021flq,Carrasco-Hidalgo:2021dyg,Contreras:2021xkf,Estrada:2018vrl,Estrada:2019aeh,daRocha:2021sqd,daRocha:2021aww,daRocha:2020jdj,daRocha:2020rda,LasHeras:2022pyj,Heras:2021xxz,LasHeras:2019wfd,Heras:2018cpz,Zubair:2022ysg,Azmat:2021kmv,Zubair:2021zqs,Zubair:2021lgt,Azmat:2021qig,Zubair:2020lna,Leon:2023nbj} (and references therein). 

This work has two main objectives. The first one, is to implement GD by means of MGD, to minimally deform the well--known exterior Schwarzschild space--time, obtaining in this way a new ``\emph{hairy}'' black hole solution. Therefore, the resulting black hole can be characterized by the primary hair $l$ (a free constant parameter) and its mass $M$. The second main target, is to obtain new wormhole solutions starting from a minimally deformed black hole. Once this process is performed, the resulting space-time bears a ``\emph{dual}'' behavior, i.e., it behaves either as a black hole or wormhole depending on the role played by the space parameter.

The present paper is organized as follows: after this short and compact introduction, we review the basic ingredients of GD by MGD approach in a four--dimensional space--time subject to spherical symmetry in Section \eqref{MGD}. Then, in Section \eqref{BHstoWH} in general how the causal structure of the original space--time is affected after MGD and also the general lines to convert the minimally deformed black hole into wormhole space-time. Subsequently, we discuss the main features of our new solution in Section \eqref{Disc}. Finally, we conclude on the findings of the work in Section \eqref{conclu}.

Along this manuscript, we shall use the most negative signature $\{+, -, -, -\}$ and relativistic geometrized units where $G=c=1$. Then the overall coupling constant $\kappa^{2}$ in front of the Einstein-Hilbert action is equal to $8\pi$ only.  

\section{The MGD protocol}\label{MGD}
This Section is devoted to introducing the main ingredients (equations as well as basic formalism) to be used in the context of the well--known GD by MGD approach \cite{Ovalle:2017fgl,Ovalle:2017wqi}. Thus, we shall review the classical Einstein's field equations and, subsequently, we shall move to the GD by means of MGD technique.
%

\subsection{Einstein equations}\label{s2}
In canonical coordinates, the space--time geometry of a spherically symmetric and static manifold, is given by
\begin{equation}
ds^{2}=
e^{\nu(r)}\,dt^{2}-e^{\lambda(r) }\,dr^{2}
-r^{2}\left( d\theta^{2}+\sin ^{2}\theta \,d\phi ^{2}\right).
\label{metric}
\end{equation}

Taking into account this general representation (\ref{metric}) and the Einstein field equations
\begin{align}
\label{EinEqFull}
G_{\mu\nu}\equiv R_{\mu \nu} - \frac{1}{2}g_{\mu \nu} R = -8\pi {T}_{\mu \nu},
\end{align}
one gets the following coupled system of equations
\begin{align}
\label{ec1}
8\pi \rho(r) &= \frac{1}{r^2}-e^{-\lambda(r)}\left[\frac{1}{r^2}-\frac{\lambda'(r)}{r}\right]\,,
\\\label{ec2}
8\pi {p}_{r}(r) &=- \frac{1}{r^2}+e^{-\lambda(r)}\left[\frac{1}{r^2}+\frac{\nu'(r)}{r}\right]\,,
\\\label{ec3}
8\pi {p}_{\perp}(r) &=\frac{1}{4}e^{-\lambda(r)}\bigg[2 \nu ''(r) + \nu'^{2}(r)-\lambda ' (r)\nu'(r) \\& \nonumber~~~
+ 2 \frac{\nu'(r)-\lambda'(r)}{r}\bigg].
\end{align}

On the left--hand side of Eqs. (\ref{ec1})--(\ref{ec3}), the quantities $\{{\rho}(r); {p}_{r}(r); {p}_{\perp}(r)\}$ 
refer to the density and to the radial and transverse pressures, respectively. 

Please note that we are considering one of the simplest cases, specifically when the energy-momentum tensor is characterized as $T_{\mu \nu} \equiv \text{diag} \{ \rho(r), {p}_{r}(r), {p}_{\perp}(r), {p}_{\perp}(r) \}$.
In more intricate scenarios, the anisotropic nature of certain sources must be accounted for by introducing an additional parameter: viscosity. For simplicity, our analysis will be confined to an anisotropic fluid, albeit with no contribution from viscosity.

The energy--momentum tensor satisfies the following conservation equation
\begin{align}\label{ConsEqEffective}
\nabla_{\mu}\,{T}^{\mu\nu}&=0\Rightarrow \frac{d{p}_{r}(r)}{dr}+\frac{\nu'(r)}{2}\bigg[{\rho}(r)+{p}_{r}(r)\bigg]-\frac{2}{r}\bigg[{p}_{\perp}(r)\\& \nonumber~~~~~~~~~~~~~-{p}_{r}(r)\bigg]=0.
\end{align}

In the framework of GD, we assume that the energy--momentum tensor can be written
as a linear combination of different fluids, namely

\begin{align}\label{StressTensorEffective}
{T}_{\mu \nu} \equiv \tilde{T}_{\mu \nu} + \alpha \theta_{\mu \nu},
\end{align}
where $\tilde{T}_{\mu \nu}$ is a known source, the so--called seed energy--momentum tensor, and $\theta_{\mu\nu}$ is a certain unknown generic fluid, the so--called decoupling fluid.

The components of such a fluid are labeled as $\theta_{\mu}^{\nu} = \text{diag}\{ \theta_{0}^{0}, \theta_{1}^{1}, \theta_{2}^{2}, \theta_{3}^{3} \}$.

Please observe that the parameter $\alpha$ is introduced artificially to regulate the impact of the anisotropic source on the isotropic background.

It is worth mentioning that, the above set of Eqs. (\ref{ec1})--(\ref{ec3}) is described in terms of effective thermodynamic quantities, encoded in the energy--momentum tensor $T_{\mu\nu}$. 

Thus, on the one hand, the effective energy-momentum defined by Eq. (\ref{StressTensorEffective}) contains six unknown functions: 
i) three coming from $\tilde{T}_{\mu\nu}$, i.e.,  $\{ \rho(r), p_r(r), p_{\perp}(r) \}$, and 
ii) three coming from $\theta_{\mu\nu}$, i.e.,  $\{ \theta_0^0(r), \theta_1^1(r), \theta_{2}^{2}(r) \}$.
On the other hand, we have two metric potentials
$\{\nu(r), \lambda(r)\}$
coming from the gravitational part, completing the eight unknown quantities.

In this case, the inclusion of the $\theta$--sector in principle introduces mathematical complications in the resolution of the system (\ref{ec1})--(\ref{ec3}) by increasing the number of degrees of freedom. However, as we will clarify shortly, the inclusion of these new terms leads to new solutions to Einstein's field equations, with certain ingredients whose physical interpretation is relevant. All this things considering the seed space--time and the GD method. So, at this point it becomes evident that these equations require a supplementary condition to close the system.
%
\subsection{GD via MGD method}

As said before, the GD by means of MGD approach allows to solve the complex set of Eqs. (\ref{ec1})--(\ref{ec3}) analytically. This elegant technique opens a new window in the construction of analytic solutions of non--linear system of equations. The main aim of this technique is to split the single complex system (\ref{ec1})--(\ref{ec3}) into two different sectors, allowing to explore their corresponding solutions, where the final solution is being the combination of solutions for each individual system. The division of this complex system of equations occurs in such a way that the field equations corresponding to the so--called seed matter content and the new source $\theta_{\mu\nu}$ appear in disjoint sections. To implement this protocol, one needs to introduce the following transformation onto the metric potentials of the space--time (\ref{metric})
\begin{eqnarray}\label{expectg}
\xi(r) &\to& \nu(r)=\xi(r)+ \alpha g(r), \label{expectg1}
\\ 
\mu(r)&\to&e^{-\lambda(r)}=\mu(r)+\alpha f(r). 
\end{eqnarray}

Notice that the functions $\xi(r)$ and $\mu(r)$ are the metric potentials when the contribution $\theta_{\mu \nu}$ is vanishing. Thus, the metric in such case takes the form
\begin{align}
    ds^{2}= e^{\xi(r)}\,dt^{2} - \frac{dr^{2}}{\mu(r)}
-r^{2}\left( d\theta^{2}+\sin ^{2}\theta \,d\phi ^{2}\right).
\label{genrel}
\end{align}

It should be noted that the radial dependency of both $g(r)$ and $f(r)$ is in order to assure the spherical symmetry and staticity of the solution. The specific MGD case corresponds to the particular case where $g(r)=0$ such that $g_{tt}(r)=e^{\nu(r)}$ is always the same and all the effects coming from the decoupling--sector lie on the radial metric potential, $g_{rr}(r)=e^{\lambda(r)}$ \cite{Ovalle:2017fgl,Ovalle:2017wqi}.

Now, plugging the transformations (\ref{expectg}) into (\ref{ec1})--(\ref{ec3}) under the mentioned considerations, one gets
\begin{align}
\label{ec1pf}
8\pi\tilde{\rho}(r)
&=\frac{1}{r^2}-\frac{\mu(r)}{r^2}-\frac{\mu'(r)}{r}\ ,
\\
\label{ec2pf}
8\pi
\tilde{p}_{r}(r)
&=
-\frac 1{r^2}+\mu(r)\left[ \frac 1{r^2}+\frac{\xi'(r)}r\right]\ ,
\\
\label{ec3pf}
8\pi
\strut\displaystyle
\tilde{p}_{\perp}(r)
&=
\frac{\mu(r)}{4}\left[2\xi''(r)+\xi'^2(r)+\frac{2\xi'(r)}{r}\right]\\& \nonumber ~~~~
+\frac{\mu'(r)}{4}\left[\xi'(r)+\frac{2}{r}\right]
\ ,
\end{align}
along with the conservation equation 
\begin{equation}
\label{conpf}
\frac{d{\tilde{p}}_{r}(r)}{dr}+\frac{\xi'(r)}{2}\left[{\tilde{\rho}}(r)+{\tilde{p}}_{r}(r)\right]-\frac{2}{r}\left[{\tilde{p}}_{\perp}(r)-{\tilde{p}}_{r}(r)\right]=0,
\end{equation}
which is a linear combination of Eqs. (\ref{ec1pf})--(\ref{ec3pf}). 

On the other hand, for the second set of equations corresponding to the additional gravitational source $\theta_{\mu\nu}$ one obtains
\begin{align}
\label{ec1d}
-8\pi\,\theta_{0}^{0}(r)
&=
\strut\displaystyle\frac{f(r)}{r^2}
+\frac{f'(r)}{r}\ ,
\\
\label{ec2d}
-8\pi
\strut\displaystyle
\,\theta_{1}^{1}(r)
&= f(r)\left[\frac{1}{r^2}+\frac{\xi'(r)}{r}\right]\ ,
\\
\label{ec3d}
-8\pi
\strut\displaystyle\,\theta_{2}^{2}(r)
&=\frac{f(r)}{4}\left[2\xi''(r)+\xi'^2(r)+2\frac{\xi'(r)}{r}\right]\\& \nonumber
+\frac{f'(r)}{4}\left[\xi'(r)+\frac{2}{r}\right]
\ .
\end{align}
The conservation equation  for the $\theta$--sector explicitly reads
\begin{equation}
\label{con1d}
\left[\theta_{1}^{1}(r)\right]'-\strut\displaystyle\frac{\xi'(r)}{2}\left[\theta_{0}^{0}(r)-\theta_{1}^{1}(r)\right]-\frac{2}{r}\left[\theta_{2}^{2}(r)-\theta_{1}^{1}(r)\right] = 0.
\end{equation}

At this stage, some observations are in order:
\begin{enumerate}
    \item The Eqs. (\ref{ec1pf})--(\ref{ec3pf}) are those corresponding to the GR theory, that is, when $\alpha=0$. This system is already solved by the seed solution characterized by line element given by \eqref{genrel}
with energy--momentum $\tilde{T}_{\mu\nu}$. It should be noted that, this energy--momentum tensor can be arbitrary, that is, a perfect fluid, anisotropic fluid, charged fluid, etc. 

\item Since the set of Eqs. (\ref{ec1pf})--(\ref{ec3pf}) is closed or solved by the seed space--time, the remaining one, that is (\ref{ec1d})--(\ref{ec3d}), should be solved in order to obtain the final space--time and its matter distribution (this system appears when $\alpha\neq0$). To close the $\theta$--sector one has a wide range of possibilities, potentially leading to a solution of physical interest. Nevertheless, as it is observed, the system (\ref{ec1d})--(\ref{ec3d}) has four unknowns, namely $\{\theta^{0}_{0}(r), \theta^{1}_{1}(r), \theta^{2}_{2}(r), f(r)\}$ and only three equations. So, it is evident that extra information is necessary to close the $\theta$--sector. In this regard, as the set (\ref{ec1d})--(\ref{ec3d}) has four unknowns, one needs at least one extra condition. This condition could be for example a well--motivated relation among the $\theta$--sector components (a type of equation of state), some constraint on the space--time scalar curvature $R(r)$, to name a few.

\item The set of Eqs. (\ref{ec1d})--(\ref{ec3d}) look very similar to the standard
spherically symmetric Einstein's field equations for an ``\emph{anisotropic}'' fluid with energy--momentum tensor $\theta_{\mu}^{\nu}=\text{diag}\left(\rho^{\theta}(r)=\theta^{0}_{0}(r), p^{\theta}_{r}(r)=-\theta^{1}_{1}(r), p^{\theta}_{\perp}(r)=-\theta^{2}_{2}(r)\right)$, whose  corresponding metric would be given by
 \begin{equation}
        ds^{2}=
e^{\xi(r)}\,dt^{2}-\frac{dr^{2}}{f(r)}
-r^{2}\left( d\theta^{2}+\sin ^{2}\theta \,d\phi ^{2}\right).
    \end{equation}

However, the right hand side of Eqs. (\ref{ec1d})--(\ref{ec3d}) are not the standard expressions for the Einstein tensor components $G^{t}_{t}$ and $G^{r}_{r}$, since there is a missing $-1/r^{2}$ in both. So, these equation are referred to the literature as quasi--Einstein field equations \cite{Ovalle:2017fgl}.

\item The expression (\ref{con1d}), is a linear combination of Eqs. (\ref{ec1d})--(\ref{ec3d}). As can be seen, both sectors are separately conserved (see Eqs. (\ref{conpf}) and (\ref{con1d}). This occurs because
\begin{align}
\nabla_{\mu}{T}^{\mu\nu}&=\nabla_{\mu}\left(\tilde{T}^{\mu\nu}+\alpha\theta^{\mu\nu}\right)=0 \\& \nonumber\Rightarrow  \nabla_{\mu}\tilde{T}^{\mu\nu}=\nabla_{\mu}{\theta}^{\mu\nu}=0,
\end{align}
meaning that there is no exchange of energy between the seed energy--momentum tensor $\tilde{T}_{\mu\nu}$ and the corresponding for the extra source, $\theta_{\mu\nu}$. They only interact gravitationally, thus ensuring in this way that Bianchi's identities are satisfied.

\item Once the $\theta$--sector is solved, the resulting space--time will be 
\begin{equation}\label{FSP}
        ds^{2}=
e^{\xi(r)}\,dt^{2}-\frac{dr^{2}}{\mu(r)+\alpha f(r)}
-r^{2}\left( d\theta^{2}+\sin ^{2}\theta \,d\phi ^{2}\right),
    \end{equation}
 where the matter distribution shall be characterized by the energy--momentum tensor (\ref{StressTensorEffective}).
\end{enumerate}

\section{From black holes to wormholes}\label{BHstoWH}
To generate black hole and wormhole solutions, we shall start considering the particular line element in Schwarszchild--like coordinates subject to the condition $|g_{tt}(r)|=|g_{rr}^{-1}(r)| \equiv h(r)$, given by
 \begin{equation}\label{eqdeformed1}
    ds^{2}=h(r)dt^{2}-\frac{dr^{2}}{h(r)}-r^{2}d\Omega^{2},
\end{equation}
where from now on we shall use the usual notation 
$d\Omega^{2}$ to denote the two sphere $d\theta^{2}+\sin ^{2}\theta \,d\phi ^{2}$.

Before to solve the problem to generate new solutions, it is instructive to analyze in general how the resulting geometric structure of the seed black hole is affected by the introduction of the decoupler function $f(r)$. So, taking into account the whole explanation given above of how the method works, and how the final space--time is expressed. Then following Eq. (\ref{FSP}), the line element (\ref{eqdeformed1}) reads
 \begin{equation}\label{eqdeformed2}
    ds^{2}=h(r)dt^{2}-\frac{dr^{2}}{h(r)+\alpha f(r)}-r^{2}d\Omega^{2}.
\end{equation}

Of course, the general Schwarzschild form $|g_{tt}(r)|=|g_{rr}^{-1}(r)|$ is not longer valid, instead here we have a minimally deformed black hole solution or a non--Schwarzschild black hole where $|g_{tt}(r)|\neq|g_{rr}^{-1}(r)|$. It is clear that under the assumption $\alpha=0$ the condition $|g_{tt}(r)|=|g_{rr}^{-1}(r)|$ and the whole original properties of the seed space--time can be recovered.  

In general, to find out the critical points of the metric (event horizons), one needs to focus on the simple zeros of the deformed metric potential $g^{-1}_{rr}(r)=0$ and its singular points\footnote{This is an important point, since new simple zeros in principle could represent new horizons (event or Cauchy horizons) and potential singular points, new real singularities besides the original one.}. 

Under the condition $|g_{tt}(r)|=|g_{rr}^{-1}(r)|$, it is obvious that the critical points of one metric potential, are the critical points of the another. Nevertheless, when MGD is applied, it is clear that there are additional critical points and both metric potentials are not satisfying the previous condition, i.e., $|g_{tt}(r)|\neq|g_{rr}^{-1}(r)|$. Hence, in principle the critical points for each metric potential are different. However, one virtue of this scheme is that the original critical points, remain being critical points for the deformed metric potential. Another interesting feature of this method is that the decoupler function contains the original metric potential. In this way, both potentials have the same simple zeros and singular points (the original ones only). In next Sections, we shall clarify all these points in a detailed way. 

As was pointed out in \cite{Ovalle:2018umz}, to preserve the original causal structure of outer static observers, once MGD is performed the new critical points should be hidden by the original event horizon, if they represent an essential singularity, in order to avoid naked singularities. On the other hand, if the critical points correspond to the coordinate singularities, in order to avoid new event horizons and change in sign in the signature of space-time, these points must be discarded. To do so, we can consider, e.g., that the size of this surface is smaller than the size of the essential singularity. To further clarify this point, suppose that the original event horizon of the black hole is $r_{H}$ and the essential singularity is located at $r_{S}$, then by solving 
\begin{equation}
    h(r^{*})+\alpha f(r^{*})=0,
\end{equation}
we determine that $r^{*}_{H}$ is a coordinate singularity (a new event horizon).
On the other hand, the new potential curvature singularities can be detected, as usual, by exploring the behavior of some invariants at that point. Now, considering the case where $r^{*}_{H}< r_{S}$, then only one event horizon corresponds to the seed solution, that is $r_{H}$. In such a case, $r^{*}_{H}$ is not relevant from the physical point of view. Although it is not mandatory at all, as said before, it would be good to preserve the outer causal structure. So, in case $r^{*}_{H}$ is not discarded, one can displace it behind the original event horizon, and the former one becomes a Cauchy horizon (see below for further details). 
So, one ends at most with a black hole space-time with: 
(i) an event horizon $r_{H}$, 
(ii) Cauchy horizon $r^{*}_{H}$, 
(iii) a central singularity at $r_{S}=0$ and finally, 
(iv) an extra curvature singularity $r^{*}_{S}>r_{S}$. 
Notice that black holes with multiple horizons (and multiple singularities) are allowed beyond Einstein's gravity \cite{Gao:2017vqv},  precisely as occurs in our case. 

Under the aforementioned conditions, it is clear that a Lorentzian traversable wormhole solution is forbidden. So, instead of hiding or discarding the new critical coordinate points, we shall redefine the validity range for the radial coordinate to be $r\in [r^{*}_{H},+\infty)$, assuming that $r^{*}_{H}>r_{H}>r^{*}_{S}$. In this way, $r^{*}_{H}$ shall represent the location of the wormhole throat (see below for more details) and not an event horizon. Again, besides the temporal component $g_{tt}$ shall not nullify at $r^{*}_{H}$, so the solution shall have no infinite redshift surface, representing in principle a traversable wormhole space--time. Notwithstanding, we cannot assure at all that the deformed space-time shall be asymptotically flat. Although this last condition can be relaxed, as long as a geodesically complete space--time is obtained. 

Generally, in applying gravitational decoupling utilizing MGD, most of the original properties of the seed space-time are preserved. It does not matter whether one is working with stellar interiors, black holes or wormholes, rather what is more important is that when MGD is present in the system as a working technique, new features and interesting ingredients with physical meaning are coming out. In view of the good properties that this methodology has, it is quite relevant to determine the new functions $\{\theta_{\mu\nu}; f(r)\}$ using reasonable prescriptions endowed with pertinent physical meaning. Among all possibilities, one can relate the components of the $\theta$--sector among them, or constrain them using the initial data, i.e., the components of the seed energy-momentum tensor. These prescriptions are known as mimic constraints \cite{Ovalle:2017fgl,Ovalle:2017wqi}. In principle, this form to close the $\theta$-sector was proposed in the compact stellar distribution realm.
{The mimic constraints can also be understood if we perform a simple dimensional analysis. Motivated by that, we can introduce a suitable ansatz for the anisotropic tensor $\theta_{\mu \nu}$.
In the context of black holes,
}
this approach could lead to some trivial solutions. Of course, it depends on the selected mimic constraint and also the form of the seed energy--momentum tensor. The original mimic constraints proposed in \cite{Ovalle:2017fgl,Ovalle:2017wqi}, were
\begin{equation}\label{mimicova}
    \theta^{0}_{0}(r)=\tilde{\rho}(r)\quad \mbox{and} \quad \theta^{1}_{1}(r)=\tilde{p}_{r}(r).
\end{equation}

As the energy-momentum tensor of the exterior Schwarzschild solution is $\tilde{T}_{\mu\nu}=0$, empty space-time, then it is clear that from Eqs. (\ref{ec1d})--(\ref{ec2d}) one obtains trivial solutions, when (\ref{mimicova}) are used. Then, this approach does not work to produce new black hole solutions, unless the seed solution has a non-zero energy-momentum tensor. 

Here, as an example to implement the above outline, we shall consider as a seed space-time the well--known exterior Schwarzschild solution of GR, given by
\begin{equation}\label{SCHBH}
ds^{2}=\left(1-\frac{2M}{r}\right)dt^{2}-\left(1-\frac{2M}{r}\right)^{-1}dr^{2}-r^{2}d\Omega^{2},  
\end{equation}
being $M$ the mass of the black hole. 

To generate black hole and wormhole solutions, we are going to employ the mimic constraint procedure. However, to overcome the mentioned situation about mimic constraints, here we impose a new one. This new mimic constraint is based on the total pressure, that is
\begin{equation}\label{mimicpre}
   \frac{1}{3}\left[\tilde{p}_{r}(r)+2\tilde{p}_{\perp}(r)\right]=\frac{1}{3}\left[\theta^{1}_{1}(r)+2\theta^{2}_{2}(r)\right].
\end{equation}

As $\tilde{p}_{r}(r)=\tilde{p}_{\perp}(r)=0$ for the Schwarzschild space--time, then the above equation becomes 
\begin{equation}\label{mimic}
\theta^{1}_{1}(r)+2\theta^{2}_{2}(r)=0.
\end{equation}

The above Eq. (\ref{mimic}), provides us a first order differential equation in $f(r)$, given by
\begin{equation}\label{ecdiff1}
f'(r)+\frac{\left[2+r\left(\xi'(r)\left(4+r\xi'(r)\right)+2r\xi''(r)\right)\right]}{r\left[2+r\xi'(r)\right]}f(r)=0.
\end{equation}

From (\ref{SCHBH}), the function $\nu$ reads
{
\begin{equation}
    \xi(r)= \ln\left(1 - \frac{2M}{r}\right),
\end{equation}
}

So, plugging the above expression into (\ref{ecdiff1}), one gets the following deformation function $f(r)$
{
\begin{equation}\label{f1}
    f(r)=\left(\frac{c_1}{r-M}\right)e^{\xi(r)},
\end{equation}
}
with $c_{1}$ an integration constant with units of length. As can be seen, the function $f(r)$ is linear with respect to the Schwarzschild metric potential. Now, the fullscale space--time is given by
{
\begin{equation}\label{full}
ds^{2}=\left(1-\frac{2M}{r}\right)dt^{2}- \frac{\left(1-\frac{2M}{r}\right)^{-1}}{\left(1+\frac{l}{r-M}\right)}dr^{2}-r^{2}d\Omega^{2},
\end{equation}
}
where $l\equiv \alpha c_{1}$ is a constant with units of length.

In the following section, we are going to analyze under what conditions, the space-time (\ref{full}) describes either a black hole (BH) or wormhole (WH) solution.  

\section{Analysis and Discussion}  \label{Disc}

In this section, we are going to discuss in detail all properties of the minimally deformed Schwarzschild space--time (\ref{full}), in order to describe either a BH or WH space-time. 

\subsection{BH solution}

Let's start by discussing the general properties that the expression (\ref{full}) must have to represent a proper BH. So, to have a proper BH space--time one needs to at least demand that the causal horizon coincides with the Killing
horizon, \i.e., $r_{H}=r^{*}_{H}$ \cite{Ovalle:2018umz}. Nevertheless, in some cases, the original event horizon $r_{H}=2M$ becomes an essential (real) singularity for the deformed BH, which is not hidden inside a large horizon, but rather arising a naked singularity. 
{
In this particular case, from (\ref{full}) we recognize two features:
(i) the function $g^{-1}_{rr}$ contains the simple zeros 
from the classical solution, and
(ii) a (new) zero point directly related to the MGD protocol, i.e., 
\begin{equation}
\begin{split}
    g^{-1}_{rr}(r)=0\Rightarrow \left(1-\frac{2M}{r}\right)\left(1+\frac{l}{r-M}\right)=0 &\\
    \Rightarrow r_{H}=2M, \quad r_{1}=M-l.
    \end{split}
\end{equation}
However, be aware and notice that $g_{rr}(r)$ has two extra critical points: the first one, a trivial point located at $r=0$, and the second one located at $r_{2}=M$. 
}
Now, we shall investigate the role played by the simple zero $r_{1}$ and the critical point $r_{2}$, and also $r_{H}$, because there is a possibility that the latter one becomes a real singularity for the deformed solution. 

First of all, we examine if the above points are essential singularities. To do so, we compute some scalar invariants such as $R(r)$, $R^{\mu\nu}(r)R_{\mu\nu}(r)$, $R^{\mu\nu\gamma\delta}(r)R_{\mu\nu\gamma\delta}(r)$. So, we have
\begin{equation}\label{ricciscalar}
    R(r)=-\frac{Ml}{\left(r-M\right)^{2}r^{2}},
\end{equation}
\begin{equation}\label{squarericcitensor}
    R_{\mu\nu}(r)R^{\mu\nu}(r)=\frac{\left(5M^{2}-6Mr+3r^{2}\right)l^{2}}{2\left(r-M\right)^{4}r^{4}},
\end{equation}
\begin{equation}\label{kretchmann}
\begin{split}
   R^{\mu\nu\gamma\delta}(r)R_{\mu\nu\gamma\delta}(r)=\frac{1}{\left(r-M\right)^{4}r^{6}}\bigg[48M^{6}+6r^{4}l^{2} &\\
   -96M^{5}\left(2r+l\right)-8Mr^{3}\left(3r+5l\right)l&\\
   +8M^{4}\left(36r^{2}+41rl+6l^{2}\right)-8M^{3}r(24r^{2}&\\+49rl
   +17l^{2})+M^{
   2}r^{2}(48r^{2}+184rl&\\
  + 125l^{2})\bigg].
   \end{split}
\end{equation}

As can be appreciated from expressions (\ref{ricciscalar})--(\ref{kretchmann}), aside the original real singularity $r_{S}=0$, the point $r_{2}=r^{*}_{S}=M$ is also a real singularity whereas the original event horizon $r_{H}=2M$ and point $r_{1}$ are not. Now, the causal structure of an exterior observer is preserved if and only if $r_{H}\geq r^{*}_{H}=r_{1}$, i.e.,
\begin{equation}\label{horizoncondition}
    l \geq-M.
\end{equation}

At this point, we can distinguish three different situations, namely:
\begin{enumerate}
    \item $-M<l<0$: In this case, the BH solution (\ref{full}) has an inner horizon, Cauchy horizon (C), given by $r^{*}_{H}=r^{*}_{C}=M-l$ and an event horizon given by $r_{H}=2M$.
    \item $l>0$: In this case, the point $r^{*}_{H}$ is not physically relevant, since in this case $r^{*}_{H}<r^{*}_{S}$. Hence, the space-time (\ref{full}) only has the event horizon given by the original one $r_{H}=2M$.
    \item $l=-M$: In this case, when the condition (\ref{horizoncondition}) is saturated, i.e., when the constant $l$ depends directly on $M$, we have a degenerated event horizon at $r_{H}=r^{*}_{H}=2M$.
\end{enumerate}
\begin{figure}[H]
  \centering
  \includegraphics[width=0.48\textwidth]{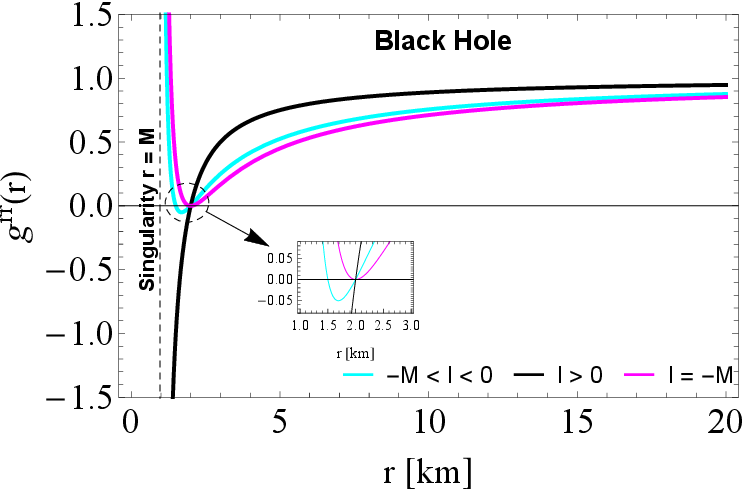} 
  \caption{The minimally deformed metric potential $g^{-1}_{rr}=g^{rr}$, for different values of the hair $l=\{-1/2;1;-1\}$ [km] (in order, magenta, black, cyan) and $M=1$ [km], against the radial coordinate $r$. }\label{fig1}
\end{figure}

Now, in considering point 1 above, it is well-known that the presence of a Cauchy horizon is associated with instabilities, what is more, the predictability of physics breaks down beyond the Cauchy horizon \cite{Hawking:1973uf}. Although we are not interested in exploring this particular problem, it is important to mention that under certain conditions imposed on the new parameter $l$, those are leading to minimally deformed BH with non--trivial causal structure inside the event horizon. On the other hand, point 3 above, said that for a minimal value of  the mass parameter $M$ compatible with $l$, both  $r_{H}$ and $r^{*}_{H}$ merge generating a degenerated event horizon at $2M$. This case could be identified as extremal--like BH as happens in the Reissner--Nordstr\"{o}m space-time when $M=Q$ \cite{Hawking:1973uf}. 

\begin{figure}[H]
  \centering
  \includegraphics[width=0.4\textwidth]{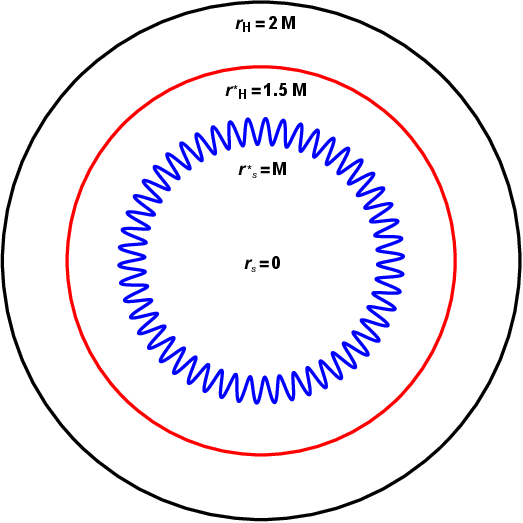} 
  \caption{ A schematic representation of the space-time (\ref{full}). Inside out: the central singularity $r_{S}=0$, the singular shell $r^{*}_{S}=M$ (blue waved circle), the Cauchy horizon $r^{*}_{C}=1.5M$ (red solid circle) and the event horizon $r_{H}=2M$ (black solid circle), for $M=1$ [km].  }\label{fig2new}
\end{figure}

So, we have obtained an asymptotically flat BH space--time (\ref{full}) characterized by the mass $M$ and a hair $l$. Nevertheless, for the extremal--like BH case, $l$ becomes a secondary hair, since it depends upon the values of the mass $M$. In Fig. \ref{fig1} the metric potential $g^{rr}(r)$ for the three cases mentioned above is shown. 
Besides, Fig. \ref{fig2new} it is displayed a schematic picture of the solution (\ref{full}), when $-M<l<0$, specifically for $l=-1/2$ [km]. In this case, the event horizon $r_{H}$ is located at $2M$ (black solid circle), the Cauchy horizon $r^{*}_{C}$ at $1.5 M$ (red solid circle), the singular shell at $r^{*}_{S}$ at $M$ (blue waved circle) and the central singularity at $r_{S}=0$.  
Notice that, for exterior solutions (black holes), one is interested in the region $r=2M$.

\begin{figure*}
  \centering
  \includegraphics[width=0.32\textwidth]{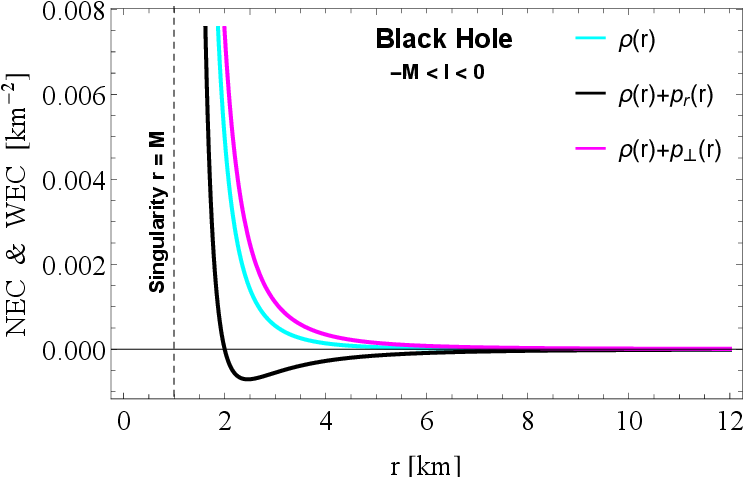} \
  \includegraphics[width=0.33\textwidth]{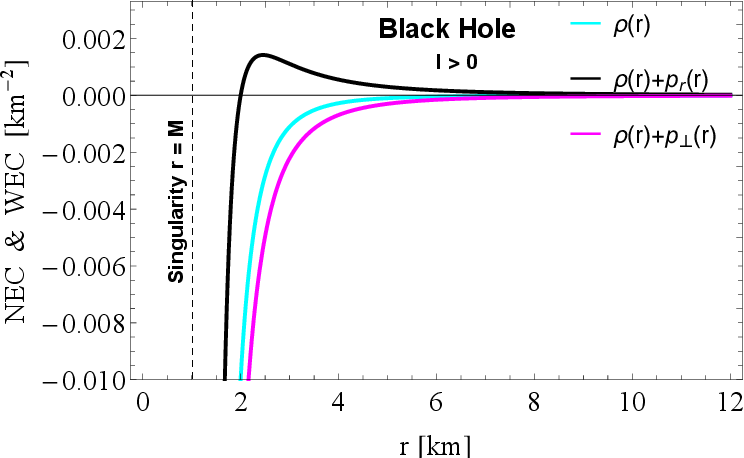} \
  \includegraphics[width=0.32\textwidth]{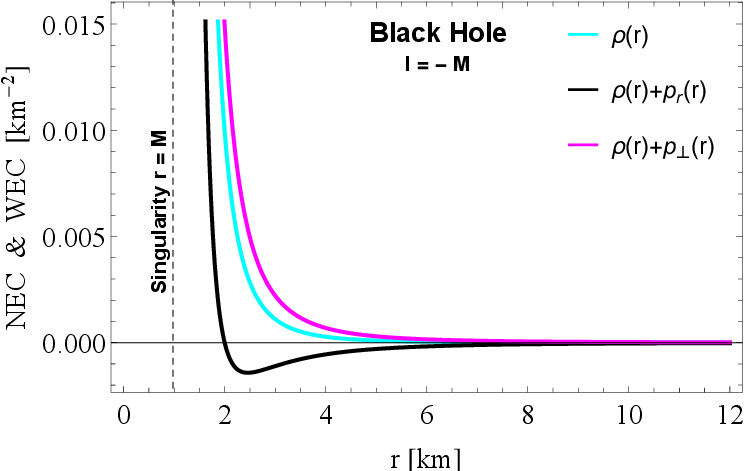}
  \caption{The null (NEC) and weak (WEC) energy conditions against the radial coordinate $r$ for the black hole solution (\ref{full}). To build these plots we considered $M=1$ [km] and $l=\{-1/2;1;-1\}$ [km] from left to right.} \label{fig2}
\end{figure*}

Next, we are going to analyze the behavior of the main thermodynamic functions to the solution (\ref{full}). So, the expressions for the density $\rho(r)$, the radial pressure $p_{r}(r)$ and transverse pressure $p_{\perp}(r)$, are given by 
\begin{eqnarray}\label{rhobh}
\rho(r)&=&\alpha \theta^{0}_{0}(r)=-\frac{Ml}{8\pi\left(r-M\right)^{2}r^{2}},\\ \label{prbh}
p_{r}(r)&=&-\alpha \theta^{1}_{1}(r)=-\frac{l}{8\pi\left(r-M\right)r^{2}},\\ \label{ptbh}
p_{\perp}(r)&=&-\alpha\theta^{2}_{2}(r)=-\frac{l}{16\pi\left(r-M\right)r^{2}}.
\end{eqnarray}

In general, it is clear from (\ref{rhobh}) that the density shall be positively defined when the hair $l$ is taking negative values (the same for the transverse pressure (\ref{ptbh})). On the other hand, the radial pressure (\ref{prbh}) shall be positive if $l<0$. In this scenario, it is not hard to see from Eqs. (\ref{rhobh}) and (\ref{prbh}), that the magnitude of the radial pressure is greater than the magnitude of the density. Therefore, it is evident that this BH solution is transgressing the so-called energy conditions (ECs). Even when $l>0$, because in this case from (\ref{rhobh}) one obtains $\rho(r)<0$. 
In this regard, as far as we know in the classical regime, states with negative density are forbidden. So, the case where the new simple zeros of the function $g^{rr}(r)$ ($r_{1}=M-l$) are not taken into account, cannot be considered as an admissible black hole solution.  

As it is well--known, the ECs are a set of restrictions imposed on the energy--momentum tensor in order to prevent nonphysical/exotic matter distribution behavior \cite{Curiel:2014zba}. 
From the classical point of view, some solutions such as wormhole space--times, are exceptions to this rule, i.e., to support these structures the presence of exotic matter violating at least the null energy condition (NEC)  is necessary \cite{Morris1988,Morris1988A,Visser:1995cc}. Moreover, in the cosmological scenario, it has been proven that the strong energy condition (SEC) is violated \cite{Visser:1997tq}. In considering the quantum regime, the violation of these conditions and the presence of states with negative density are allowed. As an example of a system violating energy conditions at the quantum level we have the well--known Casimir effect \cite{Visser:1995cc}.

In the context of black hole solutions, it becomes impossible to ensure/guarantee that all energy conditions will be satisfied (beyond the event horizon). The reason for this is due to the fact that the event horizon is a {\it{global definition}} different from the energy conditions, which are point-wise restrictions. 
Nevertheless, as pointed out in \cite{Jacobson:2007tj}, in order to have a black hole space-time satisfying the condition $g_{tt}(r)g_{rr}(r)=-1$, the energy-momentum tensor must, at least, satisfy (minimally) the NEC, i.e., the system should saturate this condition everywhere. 
In this concern, in the GR scenario, a few well-known energy-momentum tenors meet these conditions, those are: 
(i) the vacuum, 
(ii) the Maxwell field and 
(iii) cosmological constant. In this particular case, as we are filling the vacuum Schwarzschild space-time with the $\theta_{\mu\nu}$ source and deforming the radial metric potential only, it is clear that the above condition is not valid here anymore. 
To further investigate the consequences introduced by the $\theta$-sector, it is mandatory to analyze the behavior (at least) of the NEC and weak energy condition (WEC).

So, the null energy condition NEC and WEC, in general, for anisotropic matter distributions read as
\begin{equation}
    \text{NEC}: \rho(r)+p_{i}(r)\geq 0, \quad i=r,\perp,
\end{equation}
\begin{equation}
    \text{WEC}: \rho(r)\geq 0, \quad \rho(r)+p_{i}(r)\geq 0, \quad i=r,\perp.
\end{equation}

In Fig. \ref{fig2} are displayed the NEC and WEC for the space-time (\ref{full}) when it represents a BH solution. In this particular situation, in all cases shown in Fig. \ref{fig2} both, the NEC and WEC, are violated at some points. Being the worst case when $l>0$ (in this case $r^{*}_{H}$ is not relevant from the physical point of view), this is so because the density is not positively defined as it is when $r^{*}_{H}$ a Cauchy horizon (see the magenta line in the left panel of Fig. \ref{fig2}) or when $r^{*}_{H}$ coincides with $r_{H}$ (see the magenta line in the right panel of Fig. \ref{fig2}). 
However, let us reinforce that the satisfaction of energy conditions is not mandatory, and the only exception is the positiveness of the density at classical levels. Be aware and notice that when the energy conditions are satisfied we could expect a well--defined physical solution. However, one still can get well--defined solutions with physical interest when such conditions are violated. Thus, the ECs could be understood as a basic test of some solutions, but it is not conclusive.
On the other hand, if one wants to define some topological properties of black holes, such as the event horizon topology, it is necessary the satisfaction of the so--called dominant energy condition (DEC) at the event horizon location (point-wise condition). As it was proven in \cite{Hawking:1971vc}, the event horizon topology of a black hole is spherical if and only if the DEC is satisfied at this point. In general, the DEC reads as
\begin{equation}\label{dec}
    \text{DEC}: \rho(r)\geq 0, \quad -\rho(r) \leq p_{i}(r) \leq \rho(r), \quad i=r,\perp.
\end{equation}
So, taking into account the expressions (\ref{rhobh})--(\ref{ptbh}) evaluated at $r_{H}=2M$, it is easy to see that (\ref{dec}) is fulfilled for the cases $-M<l<0$ and $l=-M$, being ruled out the case $l>0$. 
In addition, when the model (\ref{full}) has inner and outer horizons and the extremal-like case, the outer most horizon (the event horizon) has spherical topology as defined in \cite{Hawking:1971vc}, contrary to what happens when $l>0$ where (\ref{dec}) is violated at $2M$. Thus, as can be appreciated under certain conditions, the presence of the $\theta$--sector introduces some topological changes in considering the event horizon shape.

\subsection{WH solution}

Taking into account the critical point $r_{1}$, instead of hiding it behind the original causal horizon $r_{H}=2M$, new structures such as WH solutions can be obtained. Obviously, the essential singularity $r_{2}$ cannot be considered. In general, to build up a WH space-time, some ingredients should be satisfied (for further details see Morris-Thorne seminal work \cite{Morris1988,Morris1988A} and Visser's book \cite{Visser:1995cc}). In Schwarzschild like--coordinates, the most general line element representing a WH structure is given by \cite{Morris1988,Morris1988A}
\begin{equation}\label{morristhorne}
    ds^{2}=e^{\Phi}dt^{2}-\frac{dr}{1-\frac{b}{r}}-r^{2}d\Omega^{2},
\end{equation}
where $\Phi=\Phi(r)$ and $b=b(r)$ are purely radial functions and, are known as the redshift and shape functions, respectively.

\begin{figure*}
  \includegraphics[width=0.31\textwidth]{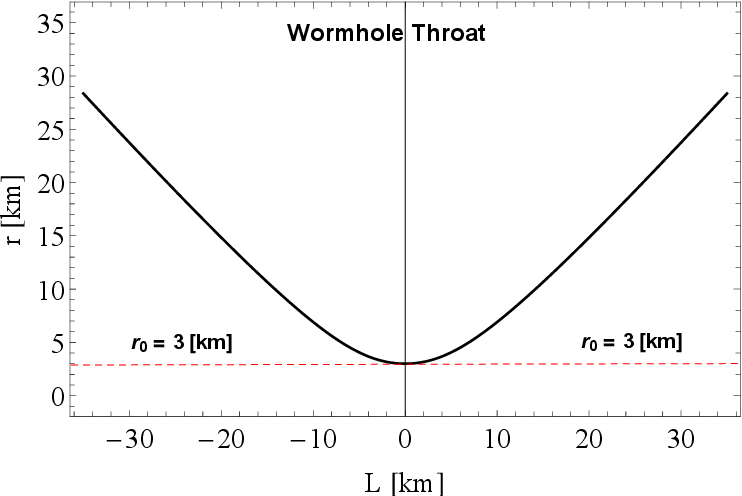} \
  \includegraphics[width=0.30\textwidth]{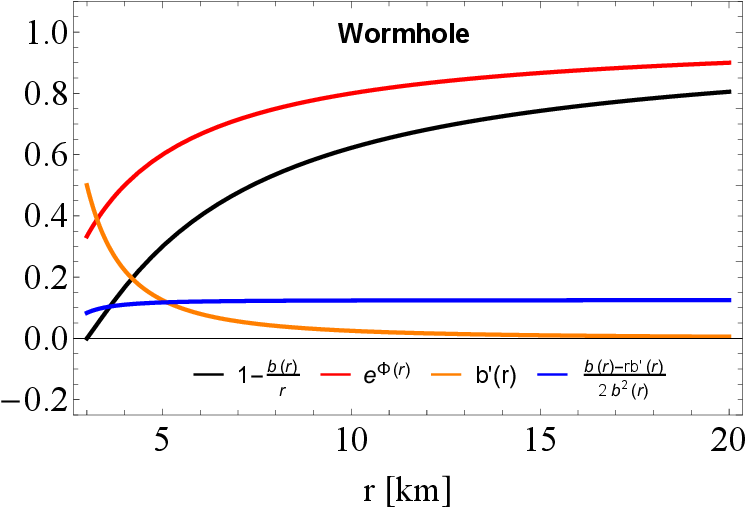} \
  \includegraphics[width=0.32\textwidth]{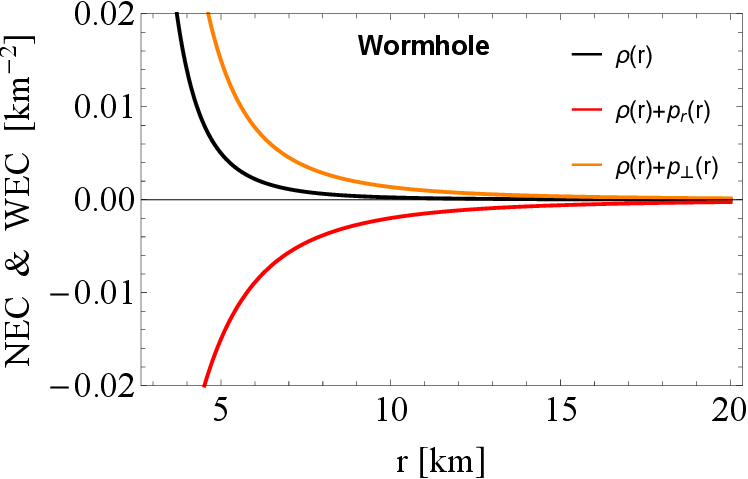}
  \caption{\textbf{Left panel}: The radial coordinate $r$ behavior against the proper radial distance $L$. The red line shows the wormhole throat location for the model (\ref{full1}) when the free parameter $l$ is equal to -2 [km]. \textbf{Middle panel}: The trend of the main geometric properties of the wormhole solution (\ref{full1}) versus the radial coordinate $r$. The inverse metric potential $g^{-1}_{rr}$ (black line), the temporal metric potential $g_{tt}$ (red line) and the flare--out condition at the throat (orange line) and beyond it (blue line). It should be noted that the vertical axis for the flare--out condition (blue line) has units of length. \textbf{Right panel}: The null (NEC) and weak (WEC) energy conditions for the model (\ref{full1}).  All these plots were obtained by considering the following space parameter: $\{M;l\}=\{1;-2\}$ [km].}\label{fig3}
\end{figure*}

Now, comparing the line elements (\ref{eqdeformed2}) and (\ref{morristhorne}), we can do the following general identification   
\begin{equation}\label{redshift}
    \Phi(r)=\ln|h(r)|,
\end{equation}
and
\begin{equation}\label{shape}
 b(r)=r\left[1-h(r)-\alpha f(r)\right].   
\end{equation}

Particularly, for the space-time (\ref{full}), one gets for the redshift function
\begin{equation}\label{red1}
    \Phi(r)=\ln\bigg{|}1-2\frac{M}{r}\bigg{|},
\end{equation}
and for the shape function
\begin{equation}\label{shape1}
  b(r)=r\left[1-{\left(1-2\frac{M}{r}\right)\left(1+\frac{l}{r-M}\right)}\right].
\end{equation}

It is evident from (\ref{shape1}), that at $r_{H}=2M$ and $r_{1}=M-l$, the shape function is equal to $r$. Therefore, both $r_{H}$ and $r_{1}$ satisfy the wormhole throat condition (see point 7 above). Nevertheless, from point 4 above, it is clear that $r_{0}=r_{H}=2M$ cannot be considered as the wormhole throat, because $\Phi(2M)=0$, leading to an infinite redshift surface.
To further corroborate this point, we need to ensure that $r_{H}=2M$ or $r_{1}=M-l$ are global maxima of the function $r=r(L)$ (see point 1). So, the candidate critical points of $dr(L)/dL$ to be maxima or minima are  
\begin{equation}\label{fderiva}
    \frac{dr(L)}{dL}=\pm\sqrt{1-\frac{b(r)}{r}}=0 \Rightarrow r_{H}=2M, \quad r_{1}=M-l.
\end{equation}

Now, computing the second derivative one has
\begin{equation}\label{secondderi}
\begin{split}
    \frac{d^{2}r(L)}{dL^{2}}=\frac{1}{2r}\left[\frac{b(r)}{r}-b'(r)\right]= & \\ \frac{1}{2r}\left(\frac{2M}{r}+
    \frac{l\left(2M-r\right)}{r\left(r-M\right)}+\frac{Ml}{\left(r-M\right)^{2}}\right).
    \end{split}
\end{equation}

So, to prove that the mentioned points are global maxima, the above expression must be strictly greater than zero at $r_{H}$ or $r_{1}$. First, we evaluate at $r_{H}=2M$ to obtain
\begin{equation}\label{bound2M}
    \frac{1}{4M}\left(1+\frac{l}{M}\right)>0 \Leftrightarrow  l>-M.
\end{equation}

Now, evaluation at $r_{1}=M-l$, we have
\begin{equation}\label{boundr1}
    \left(\frac{1}{2M-2l}\right)\left(1+\frac{M}{l}\right)>0 \Leftrightarrow \left\{ \begin{array}{lcc}
              \text{if} \quad  i)\quad  0<l<M, \\
             \\  \text{if}  \quad ii) \quad l<-M. 
             \end{array}
   \right.
\end{equation}

Remembering that the wormhole condition is $r_{1}>r_{H}$, providing $l<-M$. It is clear that (\ref{bound2M}) rules out $r_{H}$ as the wormhole throat. Furthermore, as was pointed out before, the point $r_{H}$ leads to an infinite redshift surface. Therefore, under this condition, the WH structure shall be ill-defined. So, it is evident that the wormhole throat location is at $r_{0}=r_{1}=M-l$, as shown in (\ref{boundr1}) case (ii), since (i) yields an incompatibility with the condition $r_{1}>r_{H}$. In the left panel of Fig. \ref{fig3} we have shown the behavior of the radial coordinate $r$ against the radial proper distance $L$. As can be appreciated, the radial coordinate has a global maximum when $L=0$, determining the wormhole throat location and its size ($r_{0}=3$ [km] for $l=-2$ [km]). Moreover, as $r$ moves from $r_{0}$ to $\infty$, $L$ is changing from $0$ to $\pm \infty$ as required for a WH space-time.

In summary, we have obtained a WH space-time starting from a minimally deformed BH (\ref{full}), where the new critical point $r_{1}$ plays an important role in determining the WH throat, making sure that the structure is well defined. Besides, the resulting WH has a non-trivial redshift, given by the original Shwarzschild metric potential.

Specifically, the line element of this WH solution is given by
{
\begin{equation}\label{full1}
ds^{2} = \left(1-\frac{2M}{r}\right) dt^{2} - \frac{\left(1-\frac{2M}{r}\right)^{-1}}{\left(1+\frac{l}{r-M}\right)}dr^{2} - r^{2}d\Omega^{2},
\end{equation}
}
however, the radial coordinate $r$ belongs to the interval $[r_{1},+\infty)$.

To further support the WH solution (\ref{full1}), we have corroborated the  fulfillment of some geometrical properties, necessary for the WH to exist. One of them is the so-called flare-out condition. This condition is necessary to maintain the WH open. In general, this condition reads
\begin{equation}
    \frac{b(r)-rb'(r)}{2b^{2}(r)}>0.
\end{equation}

The above expression evaluated at $r=r_{0}$ gives the following information
\begin{equation}
    b'(r_{0})<1.
\end{equation}

For the space-time (\ref{full1}), the flare-out condition at $r_{0}$ is
\begin{equation}
    -\frac{Ml}{(r_{0}-M)^{2}}<1,
\end{equation}
leading to $l<-M$ when $r_{0}=r_{1}=M-l$. 

In the middle panel of Fig. \ref{fig3} are displayed the behavior of the flare-out condition for both at $r=r_{0}$ (orange line) and for all $r>r_{0}$ (blue line). Besides, in the same panel the $g^{-1}_{rr}$ metric potential (black line) and the redshift function $\Phi(r)$ (red line) are shown. As can be seen, the inverse of the radial metric potential is everywhere positive and vanishes at the wormhole throat $r_{0}=r_{1}=3$ [km]. On the other hand, the temporal metric potential $g_{tt}=e^{\Phi(r)}$ is also positively defined and does not vanish at the throat as required to avoid an infinite redshift surface. 

Besides the mentioned features of the space-time (\ref{full1}), another relevant characteristic corresponds to the type of matter threading the structure. As it is well-known, wormhole solutions are supported by exotic matter distribution, i.e., matter violating energy conditions (at least NEC and WEC). In the context of GR theory, this condition cannot be avoided. However, as we are beyond Einstein's gravity, in principle should be possible to get a wormhole structure supported by a normal matter distribution. In this case, as we are starting from an empty space-time, all the matter sector relies on the $\theta_{\mu\nu}$ energy-momentum sector, so in principle, one can manage the output to get a normal matter distribution and at the same time the satisfaction of the flare-out condition. Here, we have obtained a WH violating both the NEC and WEC along the radial direction, although with a positively defined density (see black line in the right panel of Fig. \ref{fig3}). This resembles us to the Phantom field \cite{Lobo:2005yv,Lobo:2005us}, where for a positive density the radial pressure is negative since the equation of state (EOS) parameter obeys $\omega_{\text{Phantom}}<-1$. Moreover, the radial pressure superpass the density, thus violating the NEC and WEC. 

In our case, the (EOS) of the model is a non-linear barotropic EoS given by
\begin{equation}\label{eos}
p_{r}(\rho)=\omega_{r}\rho\Rightarrow p_{r}(\rho)=\frac{1}{2}\left(\rho-\sqrt{1+\frac{4\sqrt{2}}{\sqrt{\rho}}}\rho\right).
\end{equation}

\begin{figure}[H]
  \centering
  \includegraphics[width=0.48\textwidth]{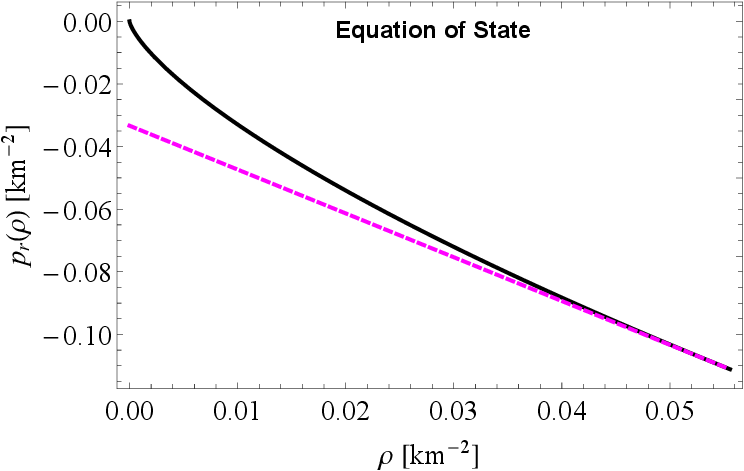} 
  \caption{The EoS for the model (\ref{full1}) (black line) and its linear approximation around the WH throat (magenta line). }\label{fig6}
\end{figure}

The trend of the radial pressure $p_{r}(\rho)$ versus the density $\rho$, is shown in the Fig. \ref{fig6} (black line). Interestingly, expanding (\ref{eos}) around the throat $r_{0}$ and keeping terms up to first order in $\rho$, the relation between $p_{r}(\rho)$ and $\rho$ is linear (see magenta line in Fig. \ref{fig6}) which can be provided as
\begin{equation}
    p_{r}(\rho)=-\frac{1}{9}-\frac{7}{5}\left(\rho-\frac{1}{18}\right)+\mathcal{O}(\rho^{2}).
\end{equation}

\begin{figure}[H]
  \centering
  \includegraphics[width=0.48\textwidth]{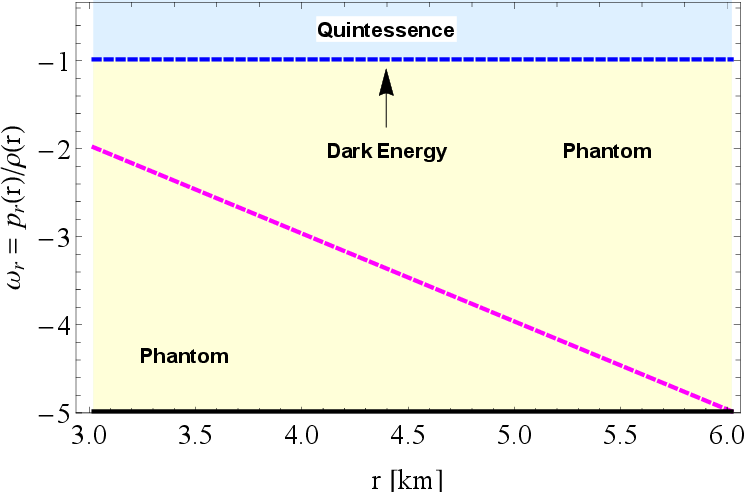} 
  \caption{The trend of the EoS parameter $\omega_{r}$ (magenta dashed line) against the radial coordinate $r$. The light-blue region represents the Quintessence regime, the dashed blue line the dark energy (or cosmological constant) and the light-yellow region the Phantom regime.  }\label{fig7}
\end{figure}

Also notice that the parameter $\omega_{r}$ in the EOS is not a constant anymore, varying with the radial coordinate $r$. Fig. \ref{fig7} is displayed its behavior against the radial coordinate $r$. As can be observed, the EOS parameter $\omega_{r}$ lies on the Phantom region. What is more, at $r_{0}$ its numerical value is -2. These sorts of WHs dominated by Phantom field with a varying EOS parameter have been studied in \cite{Cataldo:2017yec,Parsaei:2019ghw}. Nevertheless, one of the main features of our model and the one given in \cite{Cataldo:2017yec} is that in the latter one, the WH is dominated by two different kinds of exotic matter distributions. The first one is a Phantom regime and the second one is a Quintessence regime, i.e., the WH has a mixed energy dependence. In the present case, the WH solution is immersed only in a Phantom field regime. Another difference is that in \cite{Cataldo:2017yec} the WH has a finite size, whilst here the WH extends to infinity, implying that the whole space-time is filled with an exotic matter distribution. Of course, to cure this pathology we can use the well-known junction condition process \cite{Morris1988,Visser:1995cc}, by pasting our model with the outer Schwarzschild solution to confine the exotic matter in a finite region, keeping the asymptotically flat behavior and filling the space-time far away the WH with a usual matter distribution. 

To wind up this section, we want to reinforce the idea of using MGD gravitational decoupling to obtain WH solutions from minimally deformed BH driven by normal matter distributions or small amounts of exotic matter. However, this point deserves a deeper and more exhaustive analysis that shall be studied elsewhere, as well as the stability of the model.

\section{Conclusion} \label{conclu}

In the present paper, we have investigated for the first time, the transition from a minimally deformed black hole to a wormhole in the context of gravitational decoupling by means of minimally geometric deformation approach. Assuming Schwarzschild-like coordinates and a spherically symmetric background, we have obtained the effective Einstein field equations. Subsequently, with the help of a new constraint for the anisotropic sector (\ref{mimicpre}), we have obtained the decoupler function $f(r)$ (\ref{f1}), determining in this way the minimally deformed Schwarzschild space-time (\ref{full}).\\

We observe that the function $g_{rr}^{-1}(r)$ of the solution (\ref{full}) has one additional critical point $r_{1}=M-l$ and one extra real singularity $r^{*}_{S}=M$, both missing in the original GR solution, i.e., the exterior Schwarzschild space-time. To get insights regarding the singularities of our solution, the standard invariants, $R(r),~R_{\mu \nu}(r)R^{\mu \nu}(r),~R_{\mu \nu \gamma \delta}(r)R^{\mu \nu \gamma \delta}(r)$ were computed (see Eqs. (\ref{ricciscalar})--(\ref{kretchmann})), where it is clear that $r^{*}_{S}=M$ is an essential singularity. In this case, as $r^{*}_{S}>r_{S}$ (with $r_{S}=0$) 
the curvature singularity $r^{*}_{S}$ is a shell of radius $M$. This is so because, in this way, the spherical symmetry of the solution is preserved. Therefore, we obtained a BH solution with two curvature singularities.
Furthermore, the original causal structure of an external static observer is preserved iff $l \geq -M$, which implies a constraint on the parameter $l$. In principle, the parameter $l$ can be considered as a primary hair. Then the space--time (\ref{full}) is being characterized by the mass $M$ and the hair $l$. In this concern, when both surfaces $r_{H}=2M$ and $r^{*}_{H}=M-l$ coincide (in the extremal--like case), the hair $l$ becomes a secondary hair, depending on the mass $M$ parameter. So, depending on the magnitude and sign of $l$, the inner region of the BH is changing (see Fig. \ref{fig1}). On the other hand, another interesting feature of the space-time (\ref{full}) is that it preserves the asymptotically flat behavior of the original space--time. Although this is not a mandatory feature for a space--time being a BH.

Regarding the effective density (\ref{rhobh}) and pressures (\ref{prbh})--(\ref{ptbh}), their behavior depends on the sign of $l$. Subsequently, the energy conditions are investigated and, according to our results, both the NEC and WEC are violated (see Fig. \ref{fig2}). In principle, the violation of energy conditions occurs because we are beyond GR (for the Schwarzschild solution, these conditions are satisfied). Therefore, one cannot warrant their satisfaction at all unless some conditions are being imposed at the level of the field equations or a different closure for the $\theta$--sector be employed. At least, when $r^{*}_{H}$ becomes a Cauchy horizon or when $r^{*}_{H}$ merges with $r_{H}$, the density is positive defined everywhere (see left and right panels of Fig. \ref{fig2}). 
This is a remarkable feature since as far as we know, states with negative density are forbidden at the classical level. Then, it is not plausible at all to drop out or discard the possible new horizons appearing after solving the $\theta$--sector. 
Notwithstanding, as pointed out, in order to keep the original outer causal structure, at most this new horizon can be a Cauchy horizon or should be merged with the original event horizon, leading in this last case to the extremal--like BH. 

Now, moving us to the WH space-time (\ref{full1}) and performing the MGD transformation, we observe that the two critical points satisfy the WH throat condition (\ref{fderiva}). To falsify if our WH solution is well--defined, it is necessary to check which of them leads to global minima of the function $r(L)$ (\ref{secondderi}). This analysis reveals that $r^{*}_{H}$ corresponds to the WH throat $r_{0}$ (see left panel of Fig. \ref{fig3}). However, to be a proper WH structure, one needs to impose $r^{*}_{H}>r_{H}$. In this way, only $g_{rr}$ has a singular behavior at $r=r_{0}$ and not $g_{tt}$. Moreover, to be a traversable WH space--time, the flare--out condition must be satisfied. This condition is fulfilled at the throat and beyond it (see middle panel of Fig. \ref{fig3}). Besides, the NEC and WEC have been depicted in the right panel of Fig. \ref{fig3}, where it is clear that those conditions are violated as required for WH space-times. 

Last but not least, we briefly discuss the EOS of the model when it represents a WH space-time. This is a non-linear barotropic EOS (\ref{eos}). In Fig. \ref{fig6} it is displayed the trend of the EOS and in Fig. \ref{fig7} the EOS parameter $\omega_{r}$. An interesting point to be observed here is that this parameter is not a constant, what is more, it takes negative values corresponding to the Phantom regime. Therefore, the WH space--time (\ref{full1}) is driven by a Phantom field. Interestingly, the asymptotic behavior of (\ref{eos}) around $r_{0}$, shows a linear behavior. Hence, at $r_{0}$ the EOS corresponds to a linear Phantom barotropic EOS with a constant EOS parameter $\omega_{r}=-2$.

To close our conclusion part, it is pertinent to mention that there are some open questions regarding the construction of WH space-times, starting from a minimally deformed BH solution. For instance, is the space-time (\ref{full1}) stable under scalar, vector, or tensor perturbations? Is the solution (\ref{full1}) a humanely traversable WH? Since we are beyond Einstein's gravity, is it possible from a minimally deformed BH to obtain a WH that simultaneously satisfies the energy and flare-out conditions? In this respect, all these issues shall be analyzed elsewhere. However, although these questions are only of theoretical interest, once again gravitational decoupling through minimal geometric deformation has proven to be a very versatile tool to obtain new and innovative analytical solutions.

\section*{ACKNOWLEDGEMENTS}
F. Tello-Ortiz acknowledges VRIEA-PUCV
for financial support through Proyecto Postdoctorado 2023 VRIEA-PUCV. 
A. R. acknowledges financial support from the Generalitat Valenciana through PROMETEO PROJECT CIPROM/2022/13.
A. R. is funded by the María Zambrano contract ZAMBRANO 21-25 (Spain).
S. Ray gratefully acknowledges support from the Inter--University Centre for Astronomy and Astrophysics (IUCAA), Pune, India under its Visiting Research Associateship Programme as well as the facilities under ICARD, Pune at CCASS, GLA
University, Mathura, India.

\bibliography{biblio.bib}

\begin{thebibliography}{100}
\expandafter\ifx\csname url\endcsname\relax
  \def\url#1{\texttt{#1}}\fi
\expandafter\ifx\csname urlprefix\endcsname\relax\def\urlprefix{URL }\fi
\expandafter\ifx\csname href\endcsname\relax
  \def\href#1#2{#2} \def\path#1{#1}\fi

\bibitem{book:249053}
E.~P.~e. S.D. Mathur~(auth.), Physics of Black Holes: A Guided Tour, 1st
  Edition, Lecture Notes in Physics 769, Springer-Verlag Berlin Heidelberg,
  2009.

\bibitem{Bambi:2019xzp}
C.~Bambi, {Astrophysical Black Holes: A Review}, PoS MULTIF2019 (2020) 028.
\newblock \href {https://doi.org/10.22323/1.362.0028}
  {\path{doi:10.22323/1.362.0028}}.

\bibitem{Schwarzschild:1916uq}
K.~Schwarzschild, {On the gravitational field of a mass point according to
  Einstein's theory}, Sitzungsber. Preuss. Akad. Wiss. Berlin (Math. Phys. )
  1916 (1916) 189--196.
\newblock \href {http://arxiv.org/abs/physics/9905030}
  {\path{arXiv:physics/9905030}}.

\bibitem{1916AnP...355..106R}
H.~{Reissner}, {{\"U}ber die Eigengravitation des elektrischen Feldes nach der
  Einsteinschen Theorie}, Annalen der Physik 355~(9) (1916) 106--120.
\newblock \href {https://doi.org/10.1002/andp.19163550905}
  {\path{doi:10.1002/andp.19163550905}}.

\bibitem{1918KNAB...20.1238N}
G.~{Nordstr{\"o}m}, {On the Energy of the Gravitation field in Einstein's
  Theory}, Koninklijke Nederlandse Akademie van Wetenschappen Proceedings
  Series B Physical Sciences 20 (1918) 1238--1245.

\bibitem{Kerr:1963ud}
R.~P. Kerr, {Gravitational field of a spinning mass as an example of
  algebraically special metrics}, Phys. Rev. Lett. 11 (1963) 237--238.
\newblock \href {https://doi.org/10.1103/PhysRevLett.11.237}
  {\path{doi:10.1103/PhysRevLett.11.237}}.

\bibitem{Newman:1965my}
E.~T. Newman, R.~Couch, K.~Chinnapared, A.~Exton, A.~Prakash, R.~Torrence,
  {Metric of a Rotating, Charged Mass}, J. Math. Phys. 6 (1965) 918--919.
\newblock \href {https://doi.org/10.1063/1.1704351}
  {\path{doi:10.1063/1.1704351}}.

\bibitem{Hawking:1971vc}
S.~W. Hawking, {Black holes in general relativity}, Commun. Math. Phys. 25
  (1972) 152--166.
\newblock \href {https://doi.org/10.1007/BF01877517}
  {\path{doi:10.1007/BF01877517}}.

\bibitem{Ruffini:1971bza}
R.~Ruffini, J.~A. Wheeler, {Introducing the black hole}, Phys. Today 24~(1)
  (1971) 30.
\newblock \href {https://doi.org/10.1063/1.3022513}
  {\path{doi:10.1063/1.3022513}}.

\bibitem{Hawking:2016msc}
S.~W. Hawking, M.~J. Perry, A.~Strominger, {Soft Hair on Black Holes}, Phys.
  Rev. Lett. 116~(23) (2016) 231301.
\newblock \href {http://arxiv.org/abs/1601.00921} {\path{arXiv:1601.00921}},
  \href {https://doi.org/10.1103/PhysRevLett.116.231301}
  {\path{doi:10.1103/PhysRevLett.116.231301}}.

\bibitem{Kokkotas:1999bd}
K.~D. Kokkotas, B.~G. Schmidt, {Quasinormal modes of stars and black holes},
  Living Rev. Rel. 2 (1999) 2.
\newblock \href {http://arxiv.org/abs/gr-qc/9909058}
  {\path{arXiv:gr-qc/9909058}}, \href {https://doi.org/10.12942/lrr-1999-2}
  {\path{doi:10.12942/lrr-1999-2}}.

\bibitem{Berti:2009kk}
E.~Berti, V.~Cardoso, A.~O. Starinets, {Quasinormal modes of black holes and
  black branes}, Class. Quant. Grav. 26 (2009) 163001.
\newblock \href {http://arxiv.org/abs/0905.2975} {\path{arXiv:0905.2975}},
  \href {https://doi.org/10.1088/0264-9381/26/16/163001}
  {\path{doi:10.1088/0264-9381/26/16/163001}}.

\bibitem{Konoplya:2011qq}
R.~A. Konoplya, A.~Zhidenko, {Quasinormal modes of black holes: From
  astrophysics to string theory}, Rev. Mod. Phys. 83 (2011) 793--836.
\newblock \href {http://arxiv.org/abs/1102.4014} {\path{arXiv:1102.4014}},
  \href {https://doi.org/10.1103/RevModPhys.83.793}
  {\path{doi:10.1103/RevModPhys.83.793}}.

\bibitem{Panotopoulos:2017hns}
G.~Panotopoulos, A.~Rinc\'on, {Quasinormal modes of black holes in
  Einstein-power-Maxwell theory}, Int. J. Mod. Phys. D 27~(03) (2017) 1850034.
\newblock \href {http://arxiv.org/abs/1711.04146} {\path{arXiv:1711.04146}},
  \href {https://doi.org/10.1142/S0218271818500347}
  {\path{doi:10.1142/S0218271818500347}}.

\bibitem{Li:2022kch}
Z.~Li, {Scalar Perturbation Around Rotating Regular Black Hole: Superradiance
  Instability and Quasinormal Modes} (10 2022).
\newblock \href {http://arxiv.org/abs/2210.14062} {\path{arXiv:2210.14062}}.

\bibitem{Rincon:2018sgd}
A.~Rinc\'on, G.~Panotopoulos, {Quasinormal modes of scale dependent black holes
  in ( 1+2 )-dimensional Einstein-power-Maxwell theory}, Phys. Rev. D 97~(2)
  (2018) 024027.
\newblock \href {http://arxiv.org/abs/1801.03248} {\path{arXiv:1801.03248}},
  \href {https://doi.org/10.1103/PhysRevD.97.024027}
  {\path{doi:10.1103/PhysRevD.97.024027}}.

\bibitem{Destounis:2018utr}
K.~Destounis, G.~Panotopoulos, A.~Rinc\'on, {Stability under scalar
  perturbations and quasinormal modes of 4D
  Einstein\textendash{}Born\textendash{}Infeld dilaton spacetime: exact
  spectrum}, Eur. Phys. J. C 78~(2) (2018) 139.
\newblock \href {http://arxiv.org/abs/1801.08955} {\path{arXiv:1801.08955}},
  \href {https://doi.org/10.1140/epjc/s10052-018-5576-8}
  {\path{doi:10.1140/epjc/s10052-018-5576-8}}.

\bibitem{Fernando:2022wlm}
S.~Fernando, P.~A. Gonz\'alez, Y.~V\'asquez, {Extreme dilaton black holes in
  2~+~1 dimensions: quasinormal modes}, Eur. Phys. J. C 82~(7) (2022) 600.
\newblock \href {http://arxiv.org/abs/2204.02755} {\path{arXiv:2204.02755}},
  \href {https://doi.org/10.1140/epjc/s10052-022-10554-z}
  {\path{doi:10.1140/epjc/s10052-022-10554-z}}.

\bibitem{Gonzalez:2018xrq}
P.~A. Gonzalez, Y.~Vasquez, R.~N. Villalobos, {Perturbative and nonperturbative
  fermionic quasinormal modes of Einstein-Gauss-Bonnet-AdS black holes}, Phys.
  Rev. D 98~(6) (2018) 064030.
\newblock \href {http://arxiv.org/abs/1807.11827} {\path{arXiv:1807.11827}},
  \href {https://doi.org/10.1103/PhysRevD.98.064030}
  {\path{doi:10.1103/PhysRevD.98.064030}}.

\bibitem{Tzikas:2018cvs}
A.~G. Tzikas, {Bardeen black hole chemistry}, Phys. Lett. B 788 (2019)
  219--224.
\newblock \href {http://arxiv.org/abs/1811.01104} {\path{arXiv:1811.01104}},
  \href {https://doi.org/10.1016/j.physletb.2018.11.036}
  {\path{doi:10.1016/j.physletb.2018.11.036}}.

\bibitem{Bargueno:2020ais}
P.~Bargue\~no, {Some global, analytical and topological properties of regular
  black holes}, Phys. Rev. D 102~(10) (2020) 104028.
\newblock \href {http://arxiv.org/abs/2008.02680} {\path{arXiv:2008.02680}},
  \href {https://doi.org/10.1103/PhysRevD.102.104028}
  {\path{doi:10.1103/PhysRevD.102.104028}}.

\bibitem{Panotopoulos:2019qjk}
G.~Panotopoulos, A.~Rinc\'on, {Quasinormal modes of regular black holes with
  non linear-Electrodynamical sources}, Eur. Phys. J. Plus 134~(6) (2019) 300.
\newblock \href {http://arxiv.org/abs/1904.10847} {\path{arXiv:1904.10847}},
  \href {https://doi.org/10.1140/epjp/i2019-12686-x}
  {\path{doi:10.1140/epjp/i2019-12686-x}}.

\bibitem{Cruz:2004ts}
N.~Cruz, M.~Olivares, J.~R. Villanueva, {The Geodesic structure of the
  Schwarzschild anti-de Sitter black hole}, Class. Quant. Grav. 22 (2005)
  1167--1190.
\newblock \href {http://arxiv.org/abs/gr-qc/0408016}
  {\path{arXiv:gr-qc/0408016}}, \href
  {https://doi.org/10.1088/0264-9381/22/6/016}
  {\path{doi:10.1088/0264-9381/22/6/016}}.

\bibitem{Leiva:2008fd}
C.~Leiva, J.~Saavedra, J.~Villanueva, {The Geodesic Structure of the
  Schwarzschild Black Holes in Gravity's Rainbow}, Mod. Phys. Lett. A 24 (2009)
  1443--1451.
\newblock \href {http://arxiv.org/abs/0808.2601} {\path{arXiv:0808.2601}},
  \href {https://doi.org/10.1142/S0217732309029983}
  {\path{doi:10.1142/S0217732309029983}}.

\bibitem{Panotopoulos:2022bky}
G.~Panotopoulos, A.~Rincon, {Orbits of light rays in ( 1+2 )-dimensional
  Einstein-Maxwell gravity: Exact analytical solution to the null geodesic
  equations}, Ann. Phys. 443 (2022) 168947.
\newblock \href {http://arxiv.org/abs/2206.03437} {\path{arXiv:2206.03437}},
  \href {https://doi.org/10.1016/j.aop.2022.168947}
  {\path{doi:10.1016/j.aop.2022.168947}}.

\bibitem{delaCruz-Dombriz:2009pzc}
A.~de~la Cruz-Dombriz, A.~Dobado, A.~L. Maroto, {Black Holes in f(R) theories},
  Phys. Rev. D 80 (2009) 124011, [Erratum: Phys.Rev.D 83, 029903 (2011)].
\newblock \href {http://arxiv.org/abs/0907.3872} {\path{arXiv:0907.3872}},
  \href {https://doi.org/10.1103/PhysRevD.80.124011}
  {\path{doi:10.1103/PhysRevD.80.124011}}.

\bibitem{Moon:2011hq}
T.~Moon, Y.~S. Myung, E.~J. Son, {f(R) black holes}, Gen. Rel. Grav. 43 (2011)
  3079--3098.
\newblock \href {http://arxiv.org/abs/1101.1153} {\path{arXiv:1101.1153}},
  \href {https://doi.org/10.1007/s10714-011-1225-3}
  {\path{doi:10.1007/s10714-011-1225-3}}.

\bibitem{deSousaSilva:2018kkt}
M.~V. de~Sousa~Silva, M.~E. Rodrigues, {Regular black holes in $f(G)$ gravity},
  Eur. Phys. J. C 78~(8) (2018) 638.
\newblock \href {http://arxiv.org/abs/1808.05861} {\path{arXiv:1808.05861}},
  \href {https://doi.org/10.1140/epjc/s10052-018-6122-4}
  {\path{doi:10.1140/epjc/s10052-018-6122-4}}.

\bibitem{Junior:2015fya}
E.~L.~B. Junior, M.~E. Rodrigues, M.~J.~S. Houndjo, {Regular black holes in
  $f(T)$ Gravity through a nonlinear electrodynamics source}, JCAP 10 (2015)
  060.
\newblock \href {http://arxiv.org/abs/1503.07857} {\path{arXiv:1503.07857}},
  \href {https://doi.org/10.1088/1475-7516/2015/10/060}
  {\path{doi:10.1088/1475-7516/2015/10/060}}.

\bibitem{Kim:1998hc}
H.~Kim, {New black hole solutions in Brans-Dicke theory of gravity}, Phys. Rev.
  D 60 (1999) 024001.
\newblock \href {http://arxiv.org/abs/gr-qc/9811012}
  {\path{arXiv:gr-qc/9811012}}, \href
  {https://doi.org/10.1103/PhysRevD.60.024001}
  {\path{doi:10.1103/PhysRevD.60.024001}}.

\bibitem{Panotopoulos:2022xuj}
G.~Panotopoulos, A.~Rinc\'on, I.~Lopes, {Binary X-ray sources in massive
  Brans-Dicke gravity}, Universe 8 (2022) 285.
\newblock \href {http://arxiv.org/abs/2205.07412} {\path{arXiv:2205.07412}},
  \href {https://doi.org/10.3390/universe8050285}
  {\path{doi:10.3390/universe8050285}}.

\bibitem{Falls:2010he}
K.~Falls, D.~F. Litim, A.~Raghuraman, {Black Holes and Asymptotically Safe
  Gravity}, Int. J. Mod. Phys. A 27 (2012) 1250019.
\newblock \href {http://arxiv.org/abs/1002.0260} {\path{arXiv:1002.0260}},
  \href {https://doi.org/10.1142/S0217751X12500194}
  {\path{doi:10.1142/S0217751X12500194}}.

\bibitem{Cai:2010zh}
Y.-F. Cai, D.~A. Easson, {Black holes in an asymptotically safe gravity theory
  with higher derivatives}, JCAP 09 (2010) 002.
\newblock \href {http://arxiv.org/abs/1007.1317} {\path{arXiv:1007.1317}},
  \href {https://doi.org/10.1088/1475-7516/2010/09/002}
  {\path{doi:10.1088/1475-7516/2010/09/002}}.

\bibitem{Koch:2016uso}
B.~Koch, I.~A. Reyes, A.~Rinc\'on, {A scale dependent black hole in
  three-dimensional space\textendash{}time}, Class. Quant. Grav. 33~(22) (2016)
  225010.
\newblock \href {http://arxiv.org/abs/1606.04123} {\path{arXiv:1606.04123}},
  \href {https://doi.org/10.1088/0264-9381/33/22/225010}
  {\path{doi:10.1088/0264-9381/33/22/225010}}.

\bibitem{Rincon:2018lyd}
A.~Rinc\'on, B.~Koch, {Scale-dependent BTZ black hole}, Eur. Phys. J. C 78~(12)
  (2018) 1022.
\newblock \href {http://arxiv.org/abs/1806.03024} {\path{arXiv:1806.03024}},
  \href {https://doi.org/10.1140/epjc/s10052-018-6488-3}
  {\path{doi:10.1140/epjc/s10052-018-6488-3}}.

\bibitem{Rincon:2020cpz}
A.~Rinc\'on, G.~Panotopoulos, {Scale-dependent slowly rotating black holes with
  flat horizon structure}, Phys. Dark Univ. 30 (2020) 100725.
\newblock \href {http://arxiv.org/abs/2009.14678} {\path{arXiv:2009.14678}},
  \href {https://doi.org/10.1016/j.dark.2020.100725}
  {\path{doi:10.1016/j.dark.2020.100725}}.

\bibitem{Ovgun:2023ego}
A.~\"Ovg\"un, R.~C. Pantig, A.~Rinc\'on, {4D scale-dependent
  Schwarzschild-AdS/dS black holes: study of shadow and weak deflection angle
  and greybody bounding}, Eur. Phys. J. Plus 138~(3) (2023) 192.
\newblock \href {http://arxiv.org/abs/2303.01696} {\path{arXiv:2303.01696}},
  \href {https://doi.org/10.1140/epjp/s13360-023-03793-w}
  {\path{doi:10.1140/epjp/s13360-023-03793-w}}.

\bibitem{Bonanno:2000ep}
A.~Bonanno, M.~Reuter, {Renormalization group improved black hole space-times},
  Phys. Rev. D 62 (2000) 043008.
\newblock \href {http://arxiv.org/abs/hep-th/0002196}
  {\path{arXiv:hep-th/0002196}}, \href
  {https://doi.org/10.1103/PhysRevD.62.043008}
  {\path{doi:10.1103/PhysRevD.62.043008}}.

\bibitem{Ovalle:2017fgl}
J.~Ovalle, {Decoupling gravitational sources in general relativity: from
  perfect to anisotropic fluids}, Phys. Rev. D 95~(10) (2017) 104019.
\newblock \href {https://doi.org/10.1103/PhysRevD.95.104019}
  {\path{doi:10.1103/PhysRevD.95.104019}}.

\bibitem{Ovalle:2017wqi}
J.~Ovalle, R.~Casadio, R.~da~Rocha, A.~Sotomayor, {Anisotropic solutions by
  gravitational decoupling}, Eur. Phys. J. C 78~(2) (2018) 122.
\newblock \href {http://arxiv.org/abs/1708.00407} {\path{arXiv:1708.00407}},
  \href {https://doi.org/10.1140/epjc/s10052-018-5606-6}
  {\path{doi:10.1140/epjc/s10052-018-5606-6}}.

\bibitem{Ovalle:2018umz}
J.~Ovalle, R.~Casadio, R.~d. Rocha, A.~Sotomayor, Z.~Stuchlik, {Black holes by
  gravitational decoupling}, Eur. Phys. J. C 78~(11) (2018) 960.
\newblock \href {http://arxiv.org/abs/1804.03468} {\path{arXiv:1804.03468}},
  \href {https://doi.org/10.1140/epjc/s10052-018-6450-4}
  {\path{doi:10.1140/epjc/s10052-018-6450-4}}.

\bibitem{Contreras:2018nfg}
E.~Contreras, P.~Bargue\~no, {Minimal Geometric Deformation in asymptotically
  (A-)dS space-times and the isotropic sector for a polytropic black hole},
  Eur. Phys. J. C 78~(12) (2018) 985.
\newblock \href {http://arxiv.org/abs/1809.09820} {\path{arXiv:1809.09820}},
  \href {https://doi.org/10.1140/epjc/s10052-018-6472-y}
  {\path{doi:10.1140/epjc/s10052-018-6472-y}}.

\bibitem{Estrada:2020ptc}
M.~Estrada, R.~Prado, {A note of the first law of thermodynamics by
  gravitational decoupling}, Eur. Phys. J. C 80~(8) (2020) 799.
\newblock \href {http://arxiv.org/abs/2003.13168} {\path{arXiv:2003.13168}},
  \href {https://doi.org/10.1140/epjc/s10052-020-8315-x}
  {\path{doi:10.1140/epjc/s10052-020-8315-x}}.

\bibitem{Fernandes-Silva:2019fez}
A.~Fernandes-Silva, A.~J. Ferreira-Martins, R.~da~Rocha, {Extended quantum
  portrait of MGD black holes and information entropy}, Phys. Lett. B 791
  (2019) 323--330.
\newblock \href {http://arxiv.org/abs/1901.07492} {\path{arXiv:1901.07492}},
  \href {https://doi.org/10.1016/j.physletb.2019.03.010}
  {\path{doi:10.1016/j.physletb.2019.03.010}}.

\bibitem{daRocha:2020gee}
R.~a. da~Rocha, A.~A. Tomaz, {MGD-decoupled black holes, anisotropic fluids and
  holographic entanglement entropy}, Eur. Phys. J. C 80~(9) (2020) 857.
\newblock \href {http://arxiv.org/abs/2005.02980} {\path{arXiv:2005.02980}},
  \href {https://doi.org/10.1140/epjc/s10052-020-8414-8}
  {\path{doi:10.1140/epjc/s10052-020-8414-8}}.

\bibitem{Estrada:2021kuj}
M.~Estrada, {Gravitational Decoupling algorithm modifies the value of the
  conserved charges and thermodynamics properties in Lovelock Unique Vacuum
  theory}, Annals Phys. 439 (2022) 168792.
\newblock \href {http://arxiv.org/abs/2106.02166} {\path{arXiv:2106.02166}},
  \href {https://doi.org/10.1016/j.aop.2022.168792}
  {\path{doi:10.1016/j.aop.2022.168792}}.

\bibitem{Ovalle:2018gic}
J.~Ovalle, {Decoupling gravitational sources in general relativity: The
  extended case}, Phys. Lett. B 788 (2019) 213--218.
\newblock \href {http://arxiv.org/abs/1812.03000} {\path{arXiv:1812.03000}},
  \href {https://doi.org/10.1016/j.physletb.2018.11.029}
  {\path{doi:10.1016/j.physletb.2018.11.029}}.

\bibitem{Ovalle:2020kpd}
J.~Ovalle, R.~Casadio, E.~Contreras, A.~Sotomayor, {Hairy black holes by
  gravitational decoupling}, Phys. Dark Univ. 31 (2021) 100744.
\newblock \href {http://arxiv.org/abs/2006.06735} {\path{arXiv:2006.06735}},
  \href {https://doi.org/10.1016/j.dark.2020.100744}
  {\path{doi:10.1016/j.dark.2020.100744}}.

\bibitem{Contreras:2021yxe}
E.~Contreras, J.~Ovalle, R.~Casadio, {Gravitational decoupling for axially
  symmetric systems and rotating black holes}, Phys. Rev. D 103~(4) (2021)
  044020.
\newblock \href {http://arxiv.org/abs/2101.08569} {\path{arXiv:2101.08569}},
  \href {https://doi.org/10.1103/PhysRevD.103.044020}
  {\path{doi:10.1103/PhysRevD.103.044020}}.

\bibitem{Ramos:2021jta}
A.~Ramos, C.~Arias, R.~Avalos, E.~Contreras, {Geodesic motion around hairy
  black holes}, Ann. Phys. 431 (2021) 168557.
\newblock \href {http://arxiv.org/abs/2107.01146} {\path{arXiv:2107.01146}},
  \href {https://doi.org/10.1016/j.aop.2021.168557}
  {\path{doi:10.1016/j.aop.2021.168557}}.

\bibitem{Sultana:2021cvq}
J.~Sultana, {Gravitational Decoupling in Higher Order Theories}, Symmetry
  13~(9) (2021) 1598.
\newblock \href {https://doi.org/10.3390/sym13091598}
  {\path{doi:10.3390/sym13091598}}.

\bibitem{Ovalle:2021jzf}
J.~Ovalle, E.~Contreras, Z.~Stuchlik, {Kerr\textendash{}de Sitter black hole
  revisited}, Phys. Rev. D 103~(8) (2021) 084016.
\newblock \href {http://arxiv.org/abs/2104.06359} {\path{arXiv:2104.06359}},
  \href {https://doi.org/10.1103/PhysRevD.103.084016}
  {\path{doi:10.1103/PhysRevD.103.084016}}.

\bibitem{Meert:2021khi}
P.~Meert, R.~da~Rocha, {Gravitational decoupling, hairy black holes and
  conformal anomalies}, Eur. Phys. J. C 82~(2) (2022) 175.
\newblock \href {http://arxiv.org/abs/2109.06289} {\path{arXiv:2109.06289}},
  \href {https://doi.org/10.1140/epjc/s10052-022-10121-6}
  {\path{doi:10.1140/epjc/s10052-022-10121-6}}.

\bibitem{Arias:2022jax}
P.~J. Arias, P.~Bargue\~no, E.~Contreras, E.~Fuenmayor, {$2+1$
  Einstein-Klein-Gordon black holes by gravitational decoupling}, Astronomy
  1~(1) (2022) 2--14.
\newblock \href {http://arxiv.org/abs/2203.00661} {\path{arXiv:2203.00661}},
  \href {https://doi.org/10.3390/astronomy1010002}
  {\path{doi:10.3390/astronomy1010002}}.

\bibitem{Cavalcanti:2022cga}
R.~T. Cavalcanti, R.~C. de~Paiva, R.~da~Rocha, {Echoes of the gravitational
  decoupling: scalar perturbations and quasinormal modes of hairy black holes},
  Eur. Phys. J. Plus 137~(10) (2022) 1185.
\newblock \href {http://arxiv.org/abs/2203.08740} {\path{arXiv:2203.08740}},
  \href {https://doi.org/10.1140/epjp/s13360-022-03407-x}
  {\path{doi:10.1140/epjp/s13360-022-03407-x}}.

\bibitem{Cavalcanti:2022adb}
R.~T. Cavalcanti, K.~d.~S. Alves, J.~M. Hoff~da Silva, {Near-Horizon
  Thermodynamics of Hairy Black Holes from Gravitational Decoupling}, Universe
  8~(7) (2022) 363.
\newblock \href {http://arxiv.org/abs/2207.03995} {\path{arXiv:2207.03995}},
  \href {https://doi.org/10.3390/universe8070363}
  {\path{doi:10.3390/universe8070363}}.

\bibitem{Casadio:2022ndh}
R.~Casadio, A.~Giusti, J.~Ovalle, {Quantum Reissner-Nordstr\"om geometry:
  Singularity and Cauchy horizon}, Phys. Rev. D 105~(12) (2022) 124026.
\newblock \href {http://arxiv.org/abs/2203.03252} {\path{arXiv:2203.03252}},
  \href {https://doi.org/10.1103/PhysRevD.105.124026}
  {\path{doi:10.1103/PhysRevD.105.124026}}.

\bibitem{Ovalle:2022eqb}
J.~Ovalle, {Warped vacuum energy by black holes}, Eur. Phys. J. C 82~(2) (2022)
  170.
\newblock \href {http://arxiv.org/abs/2202.12037} {\path{arXiv:2202.12037}},
  \href {https://doi.org/10.1140/epjc/s10052-022-10094-6}
  {\path{doi:10.1140/epjc/s10052-022-10094-6}}.

\bibitem{Panotopoulos:2018law}
G.~Panotopoulos, A.~Rinc\'on, {Minimal Geometric Deformation in a cloud of
  strings}, Eur. Phys. J. C 78~(10) (2018) 851.
\newblock \href {http://arxiv.org/abs/1810.08830} {\path{arXiv:1810.08830}},
  \href {https://doi.org/10.1140/epjc/s10052-018-6321-z}
  {\path{doi:10.1140/epjc/s10052-018-6321-z}}.

\bibitem{Avalos:2023ywb}
R.~Avalos, P.~Bargue\~no, E.~Contreras, {A Static and Spherically Symmetric
  Hairy Black Hole in the Framework of the Gravitational Decoupling}, Fortsch.
  Phys. 71~(4-5) (2023) 2200171.
\newblock \href {http://arxiv.org/abs/2303.04119} {\path{arXiv:2303.04119}},
  \href {https://doi.org/10.1002/prop.202200171}
  {\path{doi:10.1002/prop.202200171}}.

\bibitem{Avalos:2023jeh}
R.~Avalos, E.~Contreras, {Quasi normal modes of hairy black holes at
  higher-order WKB approach}, Eur. Phys. J. C 83~(2) (2023) 155.
\newblock \href {http://arxiv.org/abs/2302.09148} {\path{arXiv:2302.09148}},
  \href {https://doi.org/10.1140/epjc/s10052-023-11288-2}
  {\path{doi:10.1140/epjc/s10052-023-11288-2}}.

\bibitem{Ovalle:2023vvu}
J.~Ovalle, {Black holes without Cauchy horizons and integrable singularities},
  Phys. Rev. D 107~(10) (2023) 104005.
\newblock \href {http://arxiv.org/abs/2305.00030} {\path{arXiv:2305.00030}},
  \href {https://doi.org/10.1103/PhysRevD.107.104005}
  {\path{doi:10.1103/PhysRevD.107.104005}}.

\bibitem{Casadio:2023iqt}
R.~Casadio, A.~Giusti, J.~Ovalle, {Quantum rotating black holes}, JHEP 05
  (2023) 118.
\newblock \href {http://arxiv.org/abs/2303.02713} {\path{arXiv:2303.02713}},
  \href {https://doi.org/10.1007/JHEP05(2023)118}
  {\path{doi:10.1007/JHEP05(2023)118}}.

\bibitem{Gabbanelli:2018bhs}
L.~Gabbanelli, A.~Rinc\'on, C.~Rubio, {Gravitational decoupled anisotropies in
  compact stars}, Eur. Phys. J. C 78~(5) (2018) 370.
\newblock \href {http://arxiv.org/abs/1802.08000} {\path{arXiv:1802.08000}},
  \href {https://doi.org/10.1140/epjc/s10052-018-5865-2}
  {\path{doi:10.1140/epjc/s10052-018-5865-2}}.

\bibitem{Ovalle:2018ans}
J.~Ovalle, R.~Casadio, R.~da~Rocha, A.~Sotomayor, Z.~Stuchlik,
  {Einstein-Klein-Gordon system by gravitational decoupling}, EPL 124~(2)
  (2018) 20004.
\newblock \href {http://arxiv.org/abs/1811.08559} {\path{arXiv:1811.08559}},
  \href {https://doi.org/10.1209/0295-5075/124/20004}
  {\path{doi:10.1209/0295-5075/124/20004}}.

\bibitem{Ovalle:2019lbs}
J.~Ovalle, C.~Posada, Z.~Stuchl\'\i{}k, {Anisotropic ultracompact Schwarzschild
  star by gravitational decoupling}, Class. Quant. Grav. 36~(20) (2019) 205010.
\newblock \href {http://arxiv.org/abs/1905.12452} {\path{arXiv:1905.12452}},
  \href {https://doi.org/10.1088/1361-6382/ab4461}
  {\path{doi:10.1088/1361-6382/ab4461}}.

\bibitem{Gabbanelli:2019txr}
L.~Gabbanelli, J.~Ovalle, A.~Sotomayor, Z.~Stuchlik, R.~Casadio, {A causal
  Schwarzschild-de Sitter interior solution by gravitational decoupling}, Eur.
  Phys. J. C 79~(6) (2019) 486.
\newblock \href {http://arxiv.org/abs/1905.10162} {\path{arXiv:1905.10162}},
  \href {https://doi.org/10.1140/epjc/s10052-019-7022-y}
  {\path{doi:10.1140/epjc/s10052-019-7022-y}}.

\bibitem{Contreras:2019iwm}
E.~Contreras, A.~Rinc\'on, P.~Bargue\~no, {A general interior anisotropic
  solution for a BTZ vacuum in the context of the Minimal Geometric Deformation
  decoupling approach}, Eur. Phys. J. C 79~(3) (2019) 216.
\newblock \href {http://arxiv.org/abs/1902.02033} {\path{arXiv:1902.02033}},
  \href {https://doi.org/10.1140/epjc/s10052-019-6749-9}
  {\path{doi:10.1140/epjc/s10052-019-6749-9}}.

\bibitem{Torres-Sanchez:2019wjv}
V.~A. Torres-S\'anchez, E.~Contreras, {Anisotropic neutron stars by
  gravitational decoupling}, Eur. Phys. J. C 79~(10) (2019) 829.
\newblock \href {http://arxiv.org/abs/1908.08194} {\path{arXiv:1908.08194}},
  \href {https://doi.org/10.1140/epjc/s10052-019-7341-z}
  {\path{doi:10.1140/epjc/s10052-019-7341-z}}.

\bibitem{Casadio:2019usg}
R.~Casadio, E.~Contreras, J.~Ovalle, A.~Sotomayor, Z.~Stuchlick,
  {Isotropization and change of complexity by gravitational decoupling}, Eur.
  Phys. J. C 79~(10) (2019) 826.
\newblock \href {http://arxiv.org/abs/1909.01902} {\path{arXiv:1909.01902}},
  \href {https://doi.org/10.1140/epjc/s10052-019-7358-3}
  {\path{doi:10.1140/epjc/s10052-019-7358-3}}.

\bibitem{Abellan:2020wjw}
G.~Abell\'an, V.~A. Torres-S\'anchez, E.~Fuenmayor, E.~Contreras, {Regularity
  condition on the anisotropy induced by gravitational decoupling in the
  framework of MGD}, Eur. Phys. J. C 80~(2) (2020) 177.
\newblock \href {http://arxiv.org/abs/2001.08573} {\path{arXiv:2001.08573}},
  \href {https://doi.org/10.1140/epjc/s10052-020-7749-5}
  {\path{doi:10.1140/epjc/s10052-020-7749-5}}.

\bibitem{Abellan:2020dze}
G.~Abell\'an, A.~Rinc\'on, E.~Fuenmayor, E.~Contreras, {Anisotropic interior
  solution by gravitational decoupling based on a non-standard anisotropy},
  Eur. Phys. J. Plus 135~(7) (2020) 606.
\newblock \href {https://doi.org/10.1140/epjp/s13360-020-00589-0}
  {\path{doi:10.1140/epjp/s13360-020-00589-0}}.

\bibitem{Contreras:2022vec}
E.~Contreras, Z.~Stuchlik, {A simple protocol to construct solutions with
  vanishing complexity by Gravitational Decoupling}, Eur. Phys. J. C 82~(8)
  (2022) 706.
\newblock \href {http://arxiv.org/abs/2208.09028} {\path{arXiv:2208.09028}},
  \href {https://doi.org/10.1140/epjc/s10052-022-10684-4}
  {\path{doi:10.1140/epjc/s10052-022-10684-4}}.

\bibitem{Santana:2022vmw}
D.~Santana, E.~Fuenmayor, E.~Contreras, {Integration of the
  Lane\textendash{}Emden equation for relativistic anisotropic polytropes
  through gravitational decoupling: a novel approach}, Eur. Phys. J. C 82~(8)
  (2022) 703.
\newblock \href {http://arxiv.org/abs/2208.09034} {\path{arXiv:2208.09034}},
  \href {https://doi.org/10.1140/epjc/s10052-022-10683-5}
  {\path{doi:10.1140/epjc/s10052-022-10683-5}}.

\bibitem{Contreras:2022nji}
E.~Contreras, Z.~Stuchlik, {Energy exchange between Tolman VII and a polytropic
  fluid}, Eur. Phys. J. C 82~(4) (2022) 365.
\newblock \href {https://doi.org/10.1140/epjc/s10052-022-10350-9}
  {\path{doi:10.1140/epjc/s10052-022-10350-9}}.

\bibitem{Ovalle:2022yjl}
J.~Ovalle, E.~Contreras, Z.~Stuchlik, {Energy exchange between relativistic
  fluids: the polytropic case}, Eur. Phys. J. C 82~(3) (2022) 211.
\newblock \href {http://arxiv.org/abs/2202.12665} {\path{arXiv:2202.12665}},
  \href {https://doi.org/10.1140/epjc/s10052-022-10168-5}
  {\path{doi:10.1140/epjc/s10052-022-10168-5}}.

\bibitem{Andrade:2021flq}
J.~Andrade, E.~Contreras, {Stellar models with like-Tolman IV complexity
  factor}, Eur. Phys. J. C 81~(10) (2021) 889.
\newblock \href {http://arxiv.org/abs/2110.10127} {\path{arXiv:2110.10127}},
  \href {https://doi.org/10.1140/epjc/s10052-021-09695-4}
  {\path{doi:10.1140/epjc/s10052-021-09695-4}}.

\bibitem{Carrasco-Hidalgo:2021dyg}
M.~Carrasco-Hidalgo, E.~Contreras, {Ultracompact stars with polynomial
  complexity by gravitational decoupling}, Eur. Phys. J. C 81~(8) (2021) 757.
\newblock \href {http://arxiv.org/abs/2108.10311} {\path{arXiv:2108.10311}},
  \href {https://doi.org/10.1140/epjc/s10052-021-09557-z}
  {\path{doi:10.1140/epjc/s10052-021-09557-z}}.

\bibitem{Contreras:2021xkf}
E.~Contreras, E.~Fuenmayor, {Gravitational cracking and complexity in the
  framework of gravitational decoupling}, Phys. Rev. D 103~(12) (2021) 124065.
\newblock \href {http://arxiv.org/abs/2107.01140} {\path{arXiv:2107.01140}},
  \href {https://doi.org/10.1103/PhysRevD.103.124065}
  {\path{doi:10.1103/PhysRevD.103.124065}}.

\bibitem{Estrada:2018vrl}
M.~Estrada, R.~Prado, {The Gravitational decoupling method: the higher
  dimensional case to find new analytic solutions}, Eur. Phys. J. Plus 134~(4)
  (2019) 168.
\newblock \href {http://arxiv.org/abs/1809.03591} {\path{arXiv:1809.03591}},
  \href {https://doi.org/10.1140/epjp/i2019-12555-8}
  {\path{doi:10.1140/epjp/i2019-12555-8}}.

\bibitem{Estrada:2019aeh}
M.~Estrada, {A way of decoupling gravitational sources in pure Lovelock
  gravity}, Eur.\, Phys.\, J.\, C 79~(11) (2019) 918, [Erratum: Eur.Phys.J.C
  80, 590 (2020)].
\newblock \href {http://arxiv.org/abs/1905.12129} {\path{arXiv:1905.12129}},
  \href {https://doi.org/10.1140/epjc/s10052-019-7444-6}
  {\path{doi:10.1140/epjc/s10052-019-7444-6}}.

\bibitem{daRocha:2021sqd}
R.~da~Rocha, {Gravitational decoupling of generalized Horndeski hybrid stars},
  Eur. Phys. J. C 82~(1) (2022) 34.
\newblock \href {http://arxiv.org/abs/2111.11995} {\path{arXiv:2111.11995}},
  \href {https://doi.org/10.1140/epjc/s10052-021-09971-3}
  {\path{doi:10.1140/epjc/s10052-021-09971-3}}.

\bibitem{daRocha:2021aww}
R.~da~Rocha, {Gravitational decoupling and superfluid stars}, Eur. Phys. J. C
  81~(9) (2021) 845.
\newblock \href {http://arxiv.org/abs/2107.13483} {\path{arXiv:2107.13483}},
  \href {https://doi.org/10.1140/epjc/s10052-021-09647-y}
  {\path{doi:10.1140/epjc/s10052-021-09647-y}}.

\bibitem{daRocha:2020jdj}
R.~da~Rocha, {Minimal geometric deformation of Yang-Mills-Dirac stellar
  configurations}, Phys. Rev. D 102~(2) (2020) 024011.
\newblock \href {http://arxiv.org/abs/2003.12852} {\path{arXiv:2003.12852}},
  \href {https://doi.org/10.1103/PhysRevD.102.024011}
  {\path{doi:10.1103/PhysRevD.102.024011}}.

\bibitem{daRocha:2020rda}
R.~a. da~Rocha, {MGD Dirac stars}, Symmetry 12~(4) (2020) 508.
\newblock \href {http://arxiv.org/abs/2002.10972} {\path{arXiv:2002.10972}},
  \href {https://doi.org/10.3390/sym12040508} {\path{doi:10.3390/sym12040508}}.

\bibitem{LasHeras:2022pyj}
C.~Las~Heras, P.~Leon, {Complexity factor of spherically anisotropic polytropes
  from gravitational decoupling}, Gen. Rel. Grav. 54~(11) (2022) 138.
\newblock \href {http://arxiv.org/abs/2203.16704} {\path{arXiv:2203.16704}},
  \href {https://doi.org/10.1007/s10714-022-03031-1}
  {\path{doi:10.1007/s10714-022-03031-1}}.

\bibitem{Heras:2021xxz}
C.~L. Heras, P.~Leon, {New interpretation of the extended geometric deformation
  in isotropic coordinates}, Eur. Phys. J. Plus 136~(8) (2021) 828.
\newblock \href {http://arxiv.org/abs/2101.09148} {\path{arXiv:2101.09148}},
  \href {https://doi.org/10.1140/epjp/s13360-021-01759-4}
  {\path{doi:10.1140/epjp/s13360-021-01759-4}}.

\bibitem{LasHeras:2019wfd}
C.~Las~Heras, P.~Le\'on, {New algorithms to obtain analytical solutions of
  Einstein\textquoteright{}s equations in isotropic coordinates}, Eur. Phys. J.
  C 79~(12) (2019) 990.
\newblock \href {http://arxiv.org/abs/1905.02380} {\path{arXiv:1905.02380}},
  \href {https://doi.org/10.1140/epjc/s10052-019-7507-8}
  {\path{doi:10.1140/epjc/s10052-019-7507-8}}.

\bibitem{Heras:2018cpz}
C.~L. Heras, P.~Leon, {Using MGD gravitational decoupling to extend the
  isotropic solutions of Einstein equations to the anisotropical domain},
  Fortsch. Phys. 66~(7) (2018) 1800036.
\newblock \href {http://arxiv.org/abs/1804.06874} {\path{arXiv:1804.06874}},
  \href {https://doi.org/10.1002/prop.201800036}
  {\path{doi:10.1002/prop.201800036}}.

\bibitem{Zubair:2022ysg}
M.~Zubair, {Stable stellar configurations with polynomial complexity factor},
  Eur. Phys. J. C 82~(11) (2022) 984.
\newblock \href {https://doi.org/10.1140/epjc/s10052-022-10959-w}
  {\path{doi:10.1140/epjc/s10052-022-10959-w}}.

\bibitem{Azmat:2021kmv}
H.~Azmat, M.~Zubair, {Anisotropic counterpart of charged Durgapal V perfect
  fluid sphere}, Int. J. Mod. Phys. D 30~(15) (2021) 2150115.
\newblock \href {https://doi.org/10.1142/S0218271821501157}
  {\path{doi:10.1142/S0218271821501157}}.

\bibitem{Zubair:2021zqs}
M.~Zubair, M.~Amin, H.~Azmat, {Anisotropic charged Heintzmann solution using
  gravitational decoupling through extended geometric deformation approach},
  Phys. Scrip. 96~(12) (2021).
\newblock \href {https://doi.org/10.1088/1402-4896/ac237d}
  {\path{doi:10.1088/1402-4896/ac237d}}.

\bibitem{Zubair:2021lgt}
M.~Zubair, H.~Azmat, M.~Amin, {Charged anisotropic fluid sphere in comparison
  with its uncharged analogue through extended geometric deformation}, Chin. J.
  Phys. 77 (2022) 898--914.
\newblock \href {https://doi.org/10.1016/j.cjph.2021.07.035}
  {\path{doi:10.1016/j.cjph.2021.07.035}}.

\bibitem{Azmat:2021qig}
H.~Azmat, M.~Zubair, {An anisotropic version of Tolman VII solution in $f(R,
  T)$ gravity via gravitational decoupling MGD approach}, Eur. Phys. J. Plus
  136~(1) (2021) 112.
\newblock \href {http://arxiv.org/abs/2106.08384} {\path{arXiv:2106.08384}},
  \href {https://doi.org/10.1140/epjp/s13360-021-01081-z}
  {\path{doi:10.1140/epjp/s13360-021-01081-z}}.

\bibitem{Zubair:2020lna}
M.~Zubair, H.~Azmat, {Anisotropic Tolman V Solution by Minimal Gravitational
  Decoupling Approach}, Ann. Phys. 420 (2020) 168248.
\newblock \href {http://arxiv.org/abs/2005.06955} {\path{arXiv:2005.06955}},
  \href {https://doi.org/10.1016/j.aop.2020.168248}
  {\path{doi:10.1016/j.aop.2020.168248}}.

\bibitem{Leon:2023nbj}
P.~Le\'on, C.~Las~Heras, {Spherically symmetric distributions with an invariant
  and vanishing complexity factor by means of the extended geometric
  deformation}, Eur. Phys. J. C 83~(3) (2023) 260.
\newblock \href {https://doi.org/10.1140/epjc/s10052-023-11415-z}
  {\path{doi:10.1140/epjc/s10052-023-11415-z}}.

\bibitem{Gao:2017vqv}
C.~Gao, Y.~Lu, S.~Yu, Y.-G. Shen, {Black hole and cosmos with multiple horizons
  and multiple singularities in vector-tensor theories}, Phys. Rev. D 97~(10)
  (2018) 104013.
\newblock \href {https://doi.org/10.1103/PhysRevD.97.104013}
  {\path{doi:10.1103/PhysRevD.97.104013}}.

\bibitem{Hawking:1973uf}
S.~W. Hawking, G.~F.~R. Ellis, {The Large Scale Structure of Space-Time},
  Cambridge Monographs on Mathematical Physics, Cambridge University Press,
  1973.
\newblock \href {https://doi.org/10.1017/CBO9780511524646}
  {\path{doi:10.1017/CBO9780511524646}}.

\bibitem{Curiel:2014zba}
E.~Curiel, {A Primer on Energy Conditions}, Einstein Stud. 13 (2017) 43--104.
\newblock \href {https://doi.org/10.1007/978-1-4939-3210-8_3}
  {\path{doi:10.1007/978-1-4939-3210-8_3}}.

\bibitem{Morris1988}
M.~S. Morris, K.~S. Thorne, Wormholes in spacetime and their use for
  interstellar travel: A tool for teaching general relativity, Am. J. Phys. 56
  (1988) 395.
\newblock \href {https://doi.org/10.1119/1.15620} {\path{doi:10.1119/1.15620}}.

\bibitem{Morris1988A}
M.~S. Morris, K.~S. Thorne, U.~Yurtsever, Wormholes, time machines, and the
  weak energy condition, Phys. Rev. Lett. 61 (1988) 1446.
\newblock \href {https://doi.org/10.1103/physrevlett.61.1446}
  {\path{doi:10.1103/physrevlett.61.1446}}.

\bibitem{Visser:1995cc}
M.~Visser, {Lorentzian wormholes: From Einstein to Hawking}, United Book Press,
  Inc., Baltimore, MD., 1995.

\bibitem{Visser:1997tq}
M.~Visser, {General relativistic energy conditions: The Hubble expansion in the
  epoch of galaxy formation}, Phys. Rev. D 56 (1997) 7578--7587.
\newblock \href {https://doi.org/10.1103/PhysRevD.56.7578}
  {\path{doi:10.1103/PhysRevD.56.7578}}.

\bibitem{Jacobson:2007tj}
T.~Jacobson, {When is g(tt) g(rr) = -1?}, Class. Quant. Grav. 24 (2007)
  5717--5719.
\newblock \href {http://arxiv.org/abs/0707.3222} {\path{arXiv:0707.3222}},
  \href {https://doi.org/10.1088/0264-9381/24/22/N02}
  {\path{doi:10.1088/0264-9381/24/22/N02}}.

\bibitem{Lobo:2005yv}
F.~S.~N. Lobo, {Stability of phantom wormholes}, Phys. Rev. D 71 (2005) 124022.
\newblock \href {https://doi.org/10.1103/PhysRevD.71.124022}
  {\path{doi:10.1103/PhysRevD.71.124022}}.

\bibitem{Lobo:2005us}
F.~S.~N. Lobo, {Phantom energy traversable wormholes}, Phys. Rev. D 71 (2005)
  084011.
\newblock \href {https://doi.org/10.1103/PhysRevD.71.084011}
  {\path{doi:10.1103/PhysRevD.71.084011}}.

\bibitem{Cataldo:2017yec}
M.~Cataldo, F.~Orellana, {Static phantom wormholes of finite size}, Phys. Rev.
  D 96~(6) (2017) 064022.
\newblock \href {https://doi.org/10.1103/PhysRevD.96.064022}
  {\path{doi:10.1103/PhysRevD.96.064022}}.

\bibitem{Parsaei:2019ghw}
F.~Parsaei, S.~Rastgoo, {Asymptotically flat wormhole solutions with variable
  equation-of-state parameter}, Phys. Rev. D 99~(10) (2019) 104037.
\newblock \href {http://arxiv.org/abs/1903.08251} {\path{arXiv:1903.08251}},
  \href {https://doi.org/10.1103/PhysRevD.99.104037}
  {\path{doi:10.1103/PhysRevD.99.104037}}.

\end{thebibliography}
\bibliographystyle{elsarticle-num}

\end{document}